# A Calculus for Collective-Adaptive Systems its Behavioural Theory[☆]

Yehia Abd Alrahman[a], Rocco De Nicola[a], Michele Loreti[b]

[a]*IMT School for Advanced Studies, Lucca, Italy*
[b]*Università di Camerino, Camerino, Italy*

**Abstract**

We propose a process calculus, named $AbC$, to study the behavioural theory of interactions in collective-adaptive systems by relying on attribute-based communication. An $AbC$ system consists of a set of parallel components each of which is equipped with a set of attributes. Communication takes place in an implicit multicast fashion, and interaction among components is dynamically established by taking into account "connections" as determined by predicates over their attributes. The structural operational semantics of $AbC$ is based on *Labeled Transition Systems* that are also used to define bisimilarity between components. Labeled bisimilarity is in full agreement with a barbed congruence, defined by relying on simple basic observables and context closure. The introduced equivalence is used to study the expressiveness of $AbC$ in terms of encoding broadcast channel-based interactions and to establish formal relationships between system descriptions at different levels of abstraction.

*Keywords:* Collective-adaptive systems, Attribute-Based Communication, Process calculus, Operational semantics, Behavioral theory

## 1. Introduction

Collective-adaptive systems (CAS) [1] are new emerging computational systems, consisting of a large number of components, featuring complex interaction mechanisms. These systems are usually distributed, heterogeneous, decentralised and interdependent, and are operating in dynamic and often unpredictable environments. CAS components combine their behaviours, by forming collectives, to achieve specific goals depending on their attributes, objectives, and functionalities. CAS are inherently scalable and their boundaries are fluid in the sense that components may enter or leave the collective at any time; so they need to dynamically adapt to their environmental conditions and contextual data. New engineering techniques to address the challenges of developing, integrating, and deploying such systems are needed [2].

Most of the current communication models and programming frameworks still handle the interaction between distributed components by relying on their identities; see, e.g., the Actor model [3], or by relying on channel-names as the case with process calculi, e.g., point-to-point [4], multicast with explicit addressing [5], or broadcast [6]. In these formalisms, interactions rely on names or addresses that are totally independent of the run-time properties, status, and capabilities of components. This makes it hard to program, coordinate, and adapt complex behaviours that highly depend on the actual status of components. Thus, a change of perspective of how communication can be derived, while possibly taking into account run-time properties, status, and capabilities of systems, is on demand.

In this article, we consider a calculus, named $AbC$, that relies on what we call *attribute-based communication*, a novel communication paradigm that permits selecting groups of partners by considering the predicates

---

[☆]This research has been partially supported by the European projects IP 257414 ASCENS and STReP 600708 QUANTICOL, and by the Italian project PRIN 2010LHT4KM CINA.

*Email addresses:* `yehia.abdalrahman@imtlucca.it` (Yehia Abd Alrahman), `rocco.denicola@imtlucca.it` (Rocco De Nicola), `michele.loreti@unicam.it` (Michele Loreti)

over the (dynamic) values of the attributes they expose. $AbC$ processes communicate anonymously in an implicit multicast fashion without a prior agreement between components; thanks to anonymity we have that dynamicity and open-endedness can be easily achieved. Interaction in $AbC$ relies on two prefixing actions:

- $(\tilde{E})@\Pi$ is the *attribute-based send* that is used to send the values of the sequence of expressions $\tilde{E}$ to those components whose attributes satisfy predicate $\Pi$;

- $\Pi(\tilde{x})$ is the *attribute-based receive* that binds to the sequence $\tilde{x}$ the values received from any component whose attributes (and possibly transmitted values) satisfy the predicate $\Pi$.

Receiving operations are blocking while sending operations are not. This breaks synchronisation dependencies between interacting partners, and permits modelling systems where communicating partners can enter or leave a group at any time without disturbing its overall behaviour. Groups are dynamically formed at the time of interaction by means of available and interested receiving components that satisfy sender's predicates. In this way, run-time changes of attributes introduce opportunistic interactions between components.

We demonstrate the expressive power of $AbC$ by showing how it can be used to encode different communication paradigms and we also provide a uniform encoding of a broadcast channel-based process calculus into $AbC$. We conjecture that the converse is not possible.

An $AbC$ system is rendered as a set of parallel components, each equipped with a set of attributes whose values can be modified by internal actions. The operational semantics of $AbC$ is given in terms of a labeled transition system (LTS) that is also used as the basis for defining a notion of bisimulation-based equivalence over $AbC$ components. We first introduce a context-based reduction barbed congruence by using very simple basic observables and then the corresponding extensional labeled bisimilarity. We show how to use the introduced bisimilarity to establish formal relationships between systems at different level of abstractions. We also prove the correctness of encoding a process calculus (inspired by CBS [6] and based on broadcast channels [7]) into $AbC$ up to our proposed equivalences.

This article is an extended and revised version of the conference paper presented in [8]. Here, we extend our behavioural theory and provide equational laws for it. Moreover, we provide full proofs of all the results introduced there. The scope of this paper is focused on the theoretical aspects of our calculus while aspects concerning programming methodologies can be found in [9]; where we show how to program complex and challenging scenarios, featuring collaboration, adaptation and reconfiguration in an intuitive way.

The rest of the paper is organised as follows. In Section 2 we formally present the syntax of $AbC$, while in Section 3 we introduce its operational semantics that is based on two relations, with the first one describing the behaviour of individual components and the second describing $AbC$ system' behaviour. In Section 4 we define a behavioural theory for $AbC$ by introducing a barbed congruence and then an equivalent definition of a labeled bisimulation. Section 5 is used to introduce a number of equational laws. In Section 6, we illustrate the expressive power of $AbC$; we discuss how the calculus can be used to model other communication paradigms and prove correctness and completeness of an encoding of a message-passing process calculus into $AbC$. Finally, in Section 7 we sum up our main contributions, relate our work to closely related state of arts and list research directions that deserve further investigation.

## 2. Syntax of the $AbC$ Calculus

In this section we formally present the syntax of $AbC$ and briefly discuss the intuition behind the different operators we introduce. We will use $\mathcal{V}$ to denote the set of *values* that can be used in an $AbC$ system. Elements in $\mathcal{V}$ are denoted by $b$, $v$ or $n$ (sometime with indexes). Moreover, we will also use the notation $\tilde{\cdot}$ to denote a sequence of *elements* and $\{\tilde{\cdot}\}$ to indicate the set of elements in the sequence.

The syntax of the $AbC$ calculus is reported in Table 1. The top-level entities of the calculus are *components* ($C$). A component, $\Gamma:_I P$, is a process $P$ associated with an *attribute environment* $\Gamma$, and an *interface* $I$. An *attribute environment* $\Gamma:\mathcal{A} \rightharpoonup \mathcal{V}$ is a partial map from attribute identifiers[1] $a \in \mathcal{A}$ to values $v \in \mathcal{V}$ where $\mathcal{A} \cap \mathcal{V} = \emptyset$. A value could be a number, a name (string), a tuple, etc. We will use Env to denote the set of *attribute environment* $\Gamma$.

---

[1] In the rest of this article, we shall occasionally use the term "attribute" instead of "attribute identifier".

An *interface* $I \subseteq \mathcal{A}$ consists of a *finite* set of *attributes* that are exposed by a component to control the interactions with other components. We will refer to the attributes in $I$ as *public attributes*, and to those in $dom(\Gamma) - I$ as *private attributes*. Components are composed using the parallel operator $\|$, e.g., $C_1 \| C_2$. To control the interactions of a component $C$, the *restriction* operators $[\ C\ ]^{\triangleleft f}$ and $[\ C\ ]^{\triangleright f}$ can be used. There $f$ is a function associating a predicate $\Pi$ to each tuple of values $\tilde{v} \in \mathcal{V}^*$ and attribute environment $\Gamma$. We will see later that $[\ C\ ]^{\triangleleft f}$ and $[\ C\ ]^{\triangleright f}$ can be used to restrict the messages that component $C$ can receive and send, respectively.

| (Components) | $C ::= \Gamma :_I P \quad \| \quad C_1 \| C_2 \quad \| \quad [\ C\ ]^{\triangleleft f} \quad \| \quad [\ C\ ]^{\triangleright f}$ |
|---|---|
| (Processes) | $P ::= \mathtt{0} \quad \| \quad \Pi(\tilde{x}).U \quad \| \quad (\tilde{E})@\Pi.U \quad \| \quad \langle\Pi\rangle P \quad \| \quad P_1 + P_2 \quad \| \quad P_1 \| P_2 \quad \| \quad K$ |
| (Updates) | $U ::= [a := E]U \quad \| \quad P$ |
| (Predicates) | $\Pi ::= \mathtt{tt} \quad \| \quad \mathtt{ff} \quad \| \quad p_k(E_1, \ldots, E_k) \quad \| \quad \Pi_1 \wedge \Pi_2 \quad \| \quad \Pi_1 \vee \Pi_2 \quad \| \quad \neg\Pi$ |
| (Expressions) | $E ::= v \quad \| \quad x \quad \| \quad a \quad \| \quad \mathtt{this}.a \quad \| \quad o_k(E_1, \ldots, E_k)$ |

Table 1: The syntax of the *AbC* calculus

A *process* $P$ can be the *inactive* process $\mathtt{0}$, an *action-prefixed* process, $act.U$, where $act$ is a communication action and $U$ is a process possibly preceded by an *attribute update*, a *context aware* process $\langle\Pi\rangle P$, a *nodeterministic choice* between two processes $P_1 + P_2$, a *parallel composition* of two processes $P_1|P_2$, or a process call with a unique identifier $K$ used in process definition $K \triangleq P$. All of these operators will now be described below. We start by explaining what we mean by expressions and predicates, then we continue by describing the actual operations on processes.

An *expression* $E$ is built from constant values $v \in \mathcal{V}$, variables $x$, attribute identifiers $a$, a reference to the value of $a$ ($\mathtt{this}.a$) in the component that is executing the code, or through a standard operators $o_k(E_1, \ldots, E_k)$. The latter indicates a generic operator with $k$-arity over *values* in $\mathcal{V}$. For the sake of simplicity, we omit the specific syntax of operators used to build expressions, we will only assume for each $o_k$ a (possibly partial) function $\mathcal{E}_{o_k} : \mathcal{V}^k \to \mathcal{V}$ describing the semantics of $o_k$. We will use $o(\tilde{E})$ when the value $k$ does not play any role or it is clear from the context. The evaluation of expression $E$ under $\Gamma$ is denoted by $[\![E]\!]_\Gamma$. The definition of $[\![\cdot]\!]_\Gamma$ is standard, the only interesting cases are $[\![a]\!]_\Gamma = [\![\mathtt{this}.a]\!]_\Gamma = \Gamma(a)$.

A *predicate* $\Pi$ is built from boolean constants, $\mathtt{tt}$ and $\mathtt{ff}$, from an *atomic predicate* $p_k(E_1, \ldots, E_k)$ and also by using standard boolean operators ($\neg$, $\wedge$ and $\vee$). The precise set of atomic predicates is not detailed here; we only assume that each $p_k$ denotes a *decidable predicates* in $\mathcal{V}^k$, i.e. $p_k \subseteq \mathcal{V}^k$. Examples of *basic predicates* are the standard binary relations like $=, >, <, \leq, \geq$.

The *satisfaction relation* $\Gamma \models \Pi$ is formally defined in Table 2 and shows when an attribute environment $\Gamma$ satisfies a predicate $\Pi$. In the rest of this paper we will use $\mathcal{M}(\Pi)$ to denote the set of *attribute environments* $\Gamma$ that satisfies $\Pi$. We also shall use the relation $\simeq$ to denote a semantic equivalence for predicates as defined below.

**Definition 2.1** (Predicate Equivalences). *Let $\Pi_1$ and $\Pi_2$ be two predicates, we have that:*

- $\Pi_1 \Rightarrow \Pi_2$ *if and only if* $\mathcal{M}(\Pi_1) \subseteq \mathcal{M}(\Pi_2)$;
- $\Pi_1 \simeq \Pi_2$ *if and only if* $\Pi_1 \Rightarrow \Pi_2$ *and* $\Pi_2 \Rightarrow \Pi_1$.

In what follows, we shall use the notation $\{\Pi\}_\Gamma$ to indicate the *closure* of predicate $\Pi$ under the attribute environment $\Gamma$; it yields a new predicate $\Pi'$ obtained from $\Pi$ after replacing each occurrence of $\mathtt{this}.a$ with its value $\Gamma(a)$. Note that, attribute identifiers occurring in an *equality operator* will also occur in its closure, e.g., for an attribute identifier $a$, we have that $\{a = v\}_\Gamma$ is equivalent to $(a = v)$ while $\{a = this.a\}_\Gamma$ is equivalent to $(a = v)$ given that $\Gamma(a) = v$.

$$\Gamma \models \text{tt for all } \Gamma$$
$$\Gamma \not\models \text{ff for all } \Gamma$$
$$\Gamma \models p_k(E_1, \ldots, E_k) \text{ iff } (\llbracket E_1 \rrbracket_\Gamma, \ldots, \llbracket E_k \rrbracket_\Gamma) \in p_k$$
$$\Gamma \models \Pi_1 \wedge \Pi_2 \text{ iff } \Gamma \models \Pi_1 \text{ and } \Gamma \models \Pi_2$$
$$\Gamma \models \Pi_1 \vee \Pi_2 \text{ iff } \Gamma \models \Pi_1 \text{ or } \Gamma \models \Pi_2$$
$$\Gamma \models \neg\Pi \text{ iff not } \Gamma \models \Pi$$

Table 2: The predicate satisfaction

The *attribute-based output* $(\tilde{E})@\Pi$ is used to send the evaluation of the sequence of expressions $\tilde{E}$ to the components whose attributes satisfy the predicate $\Pi$.

The *attribute-based input* $\Pi(\tilde{x})$ is used to receive messages from any component whose attributes (and possibly transmitted values) satisfy the predicate $\Pi$; the sequence $\tilde{x}$ acts as a placeholder for received values. Notice that the receiving predicates, used in attribute-based input actions, can also refer to variables in $\tilde{x}$ and the received values can be used to check whether specific conditions are satisfied. For instance, the action

$$((x = \text{``}try\text{''}) \wedge (\text{id} = \text{this.id} + y) \wedge (\text{this.round} = z))(x, y, z)$$

can be used to receive a message of the form $(\text{``}try\text{''}, c, r)$ where the value received on $z$ is equivalent to this.round and the value of the attribute id of the sending component is equal to this.id $+ c$. Thus, the predicate can be used to check both the received values and the values of the sending component interface.

A predicate can also refer to *local* attributes of components. Thus, an action like

$$(\text{``}try\text{''}, c, r)@(\text{partner} = \text{this.id})$$

can be used to send the message $(\text{``}try\text{''}, c, r)$ to all components whose attribute partner is equal to this.id.

An *attribute update*, $[a := E]$, is used to assign the result of the evaluation of $E$ to the attribute identifier $a$. The syntax is devised in such a way that sequences of updates are only possible after communication actions. Actually, updates can be viewed as side effects of interactions. It should be noted that the execution of a communication action and the following update(s) is atomic. This possibility allows components to modify their attribute values and thus triggering new behaviours in response to collected contextual data.

The *awareness construct*, $\langle \Pi \rangle P$, is used to trigger new behaviours (i.e., $P$) when the status of a component is changed (i.e., $\Pi \models \Gamma$). It blocks the execution of $P$ until predicate $\Pi$ is satisfied in the given attribute environment. The *parallel operator*, $P|Q$, models the interleaving between co-located processes, i.e., processes in the same component. The *choice operator*, $P + Q$, indicates a nondeterministic choice between $P$ and $Q$.

*Free and bound variables.* Input action $\Pi(\tilde{x})$ acts as binder $\tilde{x}$ in $\Pi(\tilde{x}).U$. We use $bv(P)$ and $fv(P)$ to denote the set of bound and free variables of $P$. We also assume that our processes are *closed*, i.e. without free variables ($fv(P) = \emptyset$).

## 3. AbC Operational Semantics

The operational semantics of $AbC$ is based on two relations. The transition relation $\mapsto$ that describes the behaviour of individual components and the transition relation $\rightarrow$ that relies on $\mapsto$ and describes system behaviours.

*3.1. Operational semantics of components*

We use the transition relation $\mapsto \subseteq$ Comp $\times$ CLAB $\times$ Comp to define the local behaviour of a component where Comp denotes the set of components and CLAB is the set of transition labels, $\alpha$, generated by the following grammar:

$$\alpha ::= \lambda \quad | \quad \Gamma \triangleright \widetilde{\Pi(\tilde{v})} \qquad \lambda ::= \Gamma \triangleright \overline{\Pi}(\tilde{v}) \quad | \quad \Gamma \triangleright \Pi(\tilde{v})$$

$$\frac{[\![\tilde{E}]\!]_\Gamma = \tilde{v} \quad \{\Pi_1\}_\Gamma = \Pi}{\Gamma:_I (\tilde{E})@\Pi_1.U \xrightarrow{\Gamma \downarrow I \triangleright \overline{\Pi}(\tilde{v})} \{\!| \Gamma:_I U |\!\}} \text{ BRD} \qquad \frac{}{\Gamma:_I (\tilde{E})@\Pi.P \xrightarrow{\widetilde{\Gamma' \triangleright \Pi'(\tilde{v})}} \Gamma:_I (\tilde{E})@\Pi.P} \text{ FBRD}$$

$$\frac{\Gamma' \models \{\Pi_1[\tilde{v}/\tilde{x}]\}_{\Gamma_1} \quad \Gamma_1 \downarrow I \models \Pi}{\Gamma_1:_I \Pi_1(\tilde{x}).U \xrightarrow{\Gamma' \triangleright \Pi(\tilde{v})} \{\!| \Gamma_1:_I U[\tilde{v}/\tilde{x}] |\!\}} \text{ RCV} \qquad \frac{\Gamma' \not\models \{\Pi[\tilde{v}/\tilde{x}]\}_\Gamma \vee \Gamma_1 \downarrow I \not\models \Pi'}{\Gamma_1:_I \Pi(\tilde{v}).U \xrightarrow{\widetilde{\Gamma' \triangleright \Pi'(\tilde{v})}} \Gamma_1:_I \Pi(\tilde{v}).U} \text{ FRCV}$$

$$\frac{\Gamma \models \Pi \quad \Gamma:_I P \xrightarrow{\lambda} \Gamma':_I P'}{\Gamma:_I \langle \Pi \rangle P \xrightarrow{\lambda} \Gamma':_I P'} \text{ AWARE} \qquad \frac{\Gamma \not\models \Pi}{\Gamma:_I \langle \Pi \rangle P \xrightarrow{\widetilde{\Gamma' \triangleright \Pi'(\tilde{v})}} \Gamma:_I \langle \Pi \rangle P} \text{ FAWARE1}$$

$$\frac{\Gamma \models \Pi \quad \Gamma:_I P \xrightarrow{\widetilde{\Gamma' \triangleright \Pi'(\tilde{v})}} \Gamma:_I P}{\Gamma:_I \langle \Pi \rangle P \xrightarrow{\widetilde{\Gamma' \triangleright \Pi'(\tilde{v})}} \Gamma:_I \langle \Pi \rangle P} \text{ FAWARE2}$$

Table 3: Operational Semantics of Components (Part 1)

The $\lambda$-labels are used to denote $AbC$ output $\Gamma \triangleright \overline{\Pi}(\tilde{v})$ and input $\Gamma \triangleright \Pi(\tilde{v})$ actions. The former contains the sender's predicate $\Pi$, that specifies the expected communication partners, the transmitted values $\tilde{v}$, and the portion of the sender *attribute environment* $\Gamma$ that can be perceived by receivers. The latter label is just the complementary label selected among all the possible ones that the receiver may accept.

The $\alpha$-labels include an additional label $\widetilde{\Gamma \triangleright \Pi(\tilde{v})}$ to model the case where a component is not able to receive a message. As it will be seen later, this kind of *negative* labels is crucial to appropriately handle dynamic operators like choice and awareness. In the following we will use $fn(\lambda)$ to denote the set of names occurring in $\lambda$.

The transition relation $\mapsto$ is defined in Table 3 and Table 4 inductively on the syntax of Table 1. For each process operator we have two types of rules: one describing the actions a term can perform, the other one showing how a component discards undesired input messages.

The behaviour of an *attribute-based output* is defined by rule BRD in Table 3. This rule states that when an output is executed, the sequence of expressions $\tilde{E}$ is evaluated, say to $\tilde{v}$, and the *closure* $\Pi$ of predicate $\Pi_1$ under $\Gamma$ is computed. Hence, these values are sent to other components together with $\Gamma \downarrow I$. This represents the portion of the *attribute environment* that can be perceived by the context and it is obtained from the local $\Gamma$ by limiting its domain to the attributes in the interface $I$ as defined below:

$$(\Gamma \downarrow I)(a) = \begin{cases} \Gamma(a) & a \in I \\ \bot & \text{otherwise} \end{cases}$$

Afterwards, possible updates $U$, following the action, are applied. This is expressed in terms of a recursive function $\{\!|C|\!\}$ defined below:

$$\{\!|C|\!\} = \begin{cases} \{\!| \Gamma[a \mapsto [\![E]\!]_\Gamma] :_I U |\!\} & C \equiv \Gamma:_I [a := E]U \\ \Gamma:_I P & C \equiv \Gamma:_I P \end{cases}$$

where $\Gamma[a \mapsto v]$ denotes an attribute update such that $\Gamma[a \mapsto v](a') = \Gamma(a')$ if $a \neq a'$ and $v$ otherwise. Rule BRD is not sufficient to fully describe the behaviour of an output action; we need another rule (FBRD) to model the fact that all incoming messages are *discarded* in case only output actions are possible.

Rule RCV governs the execution of input actions. It states that a message can be received when two *communication constraints* are satisfied: the local attribute environment restricted to interface $I$ ($\Gamma_1 \downarrow I$) satisfies $\Pi$, the predicate used by the sender to identify potential receivers; the sender environment $\Gamma'$ satisfies

$$\frac{\Gamma:_I P_1 \xrightarrow{\lambda} \Gamma':_I P'_1}{\Gamma:_I P_1 + P_2 \xrightarrow{\lambda} \Gamma':_I P'_1} \text{SumL} \qquad \frac{\Gamma:_I P_2 \xrightarrow{\lambda} \Gamma':_I P'_2}{\Gamma:_I P_1 + P_2 \xrightarrow{\lambda} \Gamma':_I P'_2} \text{SumR}$$

$$\frac{\Gamma:_I P_1 \xmapsto{\widetilde{\Gamma' \triangleright \Pi'(\tilde{v})}} \Gamma:_I P_1 \quad \Gamma:_I P_2 \xmapsto{\widetilde{\Gamma' \triangleright \Pi'(\tilde{v})}} \Gamma:_I P_2}{\Gamma:_I P_1 + P_2 \xmapsto{\widetilde{\Gamma' \triangleright \Pi'(\tilde{v})}} \Gamma:_I P_1 + P_2} \text{FSum}$$

$$\frac{\Gamma:_I P_1 \xrightarrow{\lambda} \Gamma':_I P'}{\Gamma:_I P_1 \mid P_2 \xrightarrow{\lambda} \Gamma':_I P' \mid P_2} \text{IntL} \qquad \frac{\Gamma:_I P_2 \xrightarrow{\lambda} \Gamma':_I P'}{\Gamma:_I P_1 \mid P_2 \xrightarrow{\lambda} \Gamma':_I P_1 \mid P'} \text{IntR}$$

$$\frac{\Gamma:_I P_1 \xmapsto{\widetilde{\Gamma' \triangleright \Pi'(\tilde{v})}} \Gamma:_I P_1 \quad \Gamma:_I P_2 \xmapsto{\widetilde{\Gamma' \triangleright \Pi'(\tilde{v})}} \Gamma:_I P_2}{\Gamma:_I P_1 \mid P_2 \xmapsto{\widetilde{\Gamma' \triangleright \Pi'(\tilde{v})}} \Gamma:_I P_1 \mid P_2} \text{FInt}$$

$$\frac{\Gamma:_I P \xrightarrow{\lambda} \Gamma':_I P' \quad K \triangleq P}{\Gamma:_I K \xrightarrow{\lambda} \Gamma':_I P'} \text{Rec} \qquad \frac{\Gamma:_I P \xmapsto{\widetilde{\Gamma' \triangleright \Pi'(\tilde{v})}} \Gamma:_I P \quad K \triangleq P}{\Gamma:_I K \xmapsto{\widetilde{\Gamma' \triangleright \Pi'(\tilde{v})}} \Gamma:_I K} \text{FRec}$$

$$\frac{}{\Gamma:_I 0 \xmapsto{\widetilde{\Gamma' \triangleright \Pi'(\tilde{v})}} \Gamma:_I 0} \text{FZero}$$

Table 4: Operational Semantics of Components (Part 2)

$$\frac{\Gamma:_I P \xmapsto{\lambda} \Gamma':_I P'}{\Gamma:_I P \xrightarrow{\lambda} \Gamma':_I P'} \text{Comp} \qquad \frac{\Gamma:_I P \xmapsto{\widetilde{\Gamma' \triangleright \Pi'(\tilde{v})}} \Gamma:_I P}{\Gamma:_I P \xrightarrow{\Gamma' \triangleright \Pi'(\tilde{v})} \Gamma:_I P} \text{FComp}$$

$$\frac{C_1 \xrightarrow{\Gamma \triangleright \Pi(\tilde{v})} C_1' \quad C_2 \xrightarrow{\Gamma \triangleright \Pi(\tilde{v})} C_2'}{C_1 \parallel C_2 \xrightarrow{\Gamma \triangleright \Pi(\tilde{v})} C_1' \parallel C_2'} \text{Sync}$$

$$\frac{C_1 \xrightarrow{\Gamma \triangleright \overline{\Pi}(\tilde{v})} C_1' \quad C_2 \xrightarrow{\Gamma \triangleright \Pi(\tilde{v})} C_2'}{C_1 \parallel C_2 \xrightarrow{\Gamma \triangleright \overline{\Pi}(\tilde{v})} C_1' \parallel C_2'} \text{ComL} \qquad \frac{C_1 \xrightarrow{\Gamma \triangleright \Pi(\tilde{v})} C_1' \quad C_2 \xrightarrow{\Gamma \triangleright \overline{\Pi}(\tilde{v})} C_2'}{C_1 \parallel C_2 \xrightarrow{\Gamma \triangleright \overline{\Pi}(\tilde{v})} C_1' \parallel C_2'} \text{ComR}$$

$$\frac{C \xrightarrow{\Gamma \triangleright \overline{\Pi}(\tilde{v})} C' \quad f(\Gamma, \tilde{v}) = \Pi'}{[\,C\,]^{\triangleright f} \xrightarrow{\Gamma \triangleright \overline{\Pi \wedge \Pi'}(\tilde{v})} [\,C'\,]^{\triangleright f}} \text{ResO} \qquad \frac{C \xrightarrow{\Gamma \triangleright \Pi \wedge \Pi'(\tilde{v})} C' \quad f(\Gamma, \tilde{v}) = \Pi'}{[\,C\,]^{\triangleleft f} \xrightarrow{\Gamma \triangleright \Pi(\tilde{v})} [\,C'\,]^{\triangleleft f}} \text{ResI}$$

$$\frac{C \xrightarrow{\Gamma \triangleright \Pi(\tilde{v})} C'}{[\,C\,]^{\triangleright f} \xrightarrow{\Gamma \triangleright \Pi(\tilde{v})} [\,C'\,]^{\triangleright f}} \text{ResOPass} \qquad \frac{C \xrightarrow{\Gamma \triangleright \overline{\Pi}(\tilde{v})} C'}{[\,C\,]^{\triangleleft f} \xrightarrow{\Gamma \triangleright \overline{\Pi}(\tilde{v})} [\,C'\,]^{\triangleleft f}} \text{ResIPass}$$

Table 5: Operational Semantics of Systems

the receiving predicate $\{\Pi_1[\tilde{v}/\tilde{x}]\}_{\Gamma_1}$. When these two constraints are satisfied the input action is performed and the update $U$ is applied under the substitution $[\tilde{v}/\tilde{x}]$.

Rule FRcv states that an input is *discarded* when the local attribute environment does not satisfy the *sender's predicate*, or the *receiving predicate* is not satisfied by the sender's environment.

The behaviour of a component $\Gamma:_I \langle \Pi \rangle P$ is the same as of $\Gamma:_I P$ only when $\Gamma \models \Pi$, while the component is inactive when $\Gamma \not\models \Pi$. This is rendered by rules Aware, FAware1 and FAware2.

Rules SumL, SumR, and FSum describe behaviour of $\Gamma:_I P_1 + P_2$. Rules SumL and SumR are standard and just say that $\Gamma:_I P_1 + P_2$ behaves nondeterministically either like $\Gamma:_I P_1$ or like $\Gamma:_I P_2$. A message is *discarded* by $\Gamma:_I P_1 + P_2$ if and only if both $P_1$ and $P_2$ are not able to receive it. We can observe here that the presence of discarding rules is fundamental to prevent processes that cannot receive messages from evolving without performing actions. Thus *dynamic operators*, that are the ones *disappearing* after a transition like awareness and choice, persist after a message refusal.

The behaviour of the interleaving operator is described by rules IntL, IntR and FInt. The first two are standard process algebraic rules for parallel composition while the discarding rule FInt has a similar interpretation as of rule FSum: a message can be discarded only if both the parallel processes can discard it.

Finally, rules Rec, FRec and FZero are the standard rules for handling process definition and the inactive process. The latter states that process **0** always discards messages.

*3.2. Operational semantics of systems*

The behaviour of an *AbC* system is described by means of the transition relation $\to \subseteq \text{Comp} \times \text{SLAB} \times \text{Comp}$, where Comp denotes the set of components and SLAB is the set of transition labels, $\lambda$, generated by the following grammar:

$$\lambda ::= \Gamma \triangleright \overline{\Pi}(\tilde{v}) \quad | \quad \Gamma \triangleright \Pi(\tilde{v})$$

The definition of the transition relation $\to$ is provided in Table 5.

Rules Comp and FComp depends on relation $\mapsto$ and they are used to lift the effect of local behaviour to the system level. The former rule states that the relations $\mapsto$ and $\to$ coincide when performing either an input or an output actions, while rule FComp states that a component $\Gamma:_I P$ can discard a message and remain unchanged. However, we would like to stress that the system level label of FComp coincides with that of Comp in case of input actions, which means that externally it cannot be observed whether a message has been accepted or discarded.

Rule Sync states that two parallel components $C_1$ and $C_2$ can receive the same message. Rule ComL and its symmetric variant ComR govern communication between two parallel components $C_1$ and $C_2$.

Rules ResO and ResI show how *restriction operators* $[\,C\,]^{\triangleright f}$ and $[\,C\,]^{\triangleleft f}$ limit *output* and *input* capabilities of $C$ under function $f$.

Rule ResO states that if $C$ evolves to $C'$ with label $\Gamma \triangleright \overline{\Pi}(\tilde{v})$ and $f(\Gamma, \tilde{v}) = \Pi'$ then $[\,C\,]^{\triangleright f}$ evolves with label $\Gamma \triangleright \overline{\Pi \wedge \Pi'}(\tilde{v})$ to $[\,C'\,]^{\triangleright f}$. This means that when $C$ sends messages to all the components satisfying $\Pi$, the *restriction operator* limits the interaction to only those that also satisfy $\Pi'$.

Rule ResI is similar. However, in this case, the *restriction operator* limits the input capabilities of $C$. Indeed, $[\,C\,]^{\triangleleft f}$ will receive the message $\tilde{v}$ and evolve to $[\,C'\,]^{\triangleleft f}$ with a label $\Gamma \triangleright \Pi(\tilde{v})$ only when $C \xrightarrow{\Gamma \triangleright \Pi \wedge \Pi'(\tilde{v})} C'$ where $f(\Gamma, \tilde{v}) = \Pi'$. Thus, message $\tilde{v}$ is delivered only to those components that satisfy both $\Pi$ and $\Pi'$. Note that, both $[\,C\,]^{\triangleright f}$ and $[\,C\,]^{\triangleleft f}$ completely hide input/output capabilities whenever $f(\Gamma, \tilde{v}) \wedge \Pi \simeq \mathsf{ff}$.

Rule ResOPass (resp. ResIPass) states any *input transition* (resp. *output transition*) performed by $C$ is also done by $[\,C\,]^{\triangleright f}$ (resp. $[\,C\,]^{\triangleleft f}$).

In what follows, we shall use the following notations:

- $C \xrightarrow{\tau} C'$ iff $\exists \tilde{v}, \Gamma,$ and $\Pi$ such that $C \xrightarrow{\Gamma \triangleright \overline{\Pi}(\tilde{v})} C'$ and $\Pi \simeq \mathsf{ff}$. In this case, we say that $\lambda = \Gamma \triangleright \overline{\Pi}(\tilde{v})$ is *silent*, and write $\lambda = \tau$. We write $\lambda \neq \tau$ to indicate that $\lambda$ is not silent.

- $\Rightarrow$ denotes $(\xrightarrow{\tau})^*$.

- $\xRightarrow{\lambda}$ denotes $\Rightarrow \xrightarrow{\lambda} \Rightarrow$ if $(\lambda \neq \tau)$.

- $\xRightarrow{\widehat{\lambda}}$ denotes $\Rightarrow$ if $(\lambda = \tau)$ and $\xRightarrow{\lambda}$ otherwise.

- $C \xhookrightarrow{\Pi} C'$ if and only if $\exists \tilde{v}, \Gamma,$ and $\Pi'$ such that $\Pi \simeq \Pi'$ and $C \xrightarrow{\Gamma \triangleright \overline{\Pi'}(\tilde{v})} C'$, while $C \xhookrightarrow{\Pi}_\tau C'$ if and only if $\exists \tilde{v}, \Gamma,$ and $\Pi'$ such that $\Pi \simeq \Pi'$ and $C \xRightarrow{\widehat{\Gamma \triangleright \overline{\Pi'}(\tilde{v})}} C'$.

**Lemma 3.1.** *For any* AbC *component, the following properties hold:*

1. *For any $\lambda$ such that $\lambda = \Gamma' \triangleright \Pi(\tilde{v})$ and $\Pi \simeq \mathsf{ff}$, then $C \xrightarrow{\lambda} C$;*
2. *if $C_1 \xrightarrow{\lambda} C_1'$ and $\lambda = \tau$, then $C_1 \| C \xrightarrow{\tau} C_1' \| C$ and $C \| C_1 \xrightarrow{\tau} C \| C_1'$;*
3. *if $C_1 \Rightarrow C_1'$ then $C_1 \| C \Rightarrow C_1' \| C$ and $C \| C_1 \Rightarrow C \| C_1'$;*
4. *if $C_1 \xrightarrow{\Gamma \triangleright \Pi_1(\tilde{v})} C_1'$ and $\Pi_1 \simeq \Pi_2$ then $C_1 \xrightarrow{\Gamma \triangleright \Pi_2(\tilde{v})} C_1'$;*
5. *if $C_1 \xrightarrow{\tau} C_1'$, then for any $f$: $[\,C_1\,]^{\triangleright f} \xrightarrow{\tau} [\,C_1'\,]^{\triangleright f}$ and $[\,C_1\,]^{\triangleleft f} \xrightarrow{\tau} [\,C_1'\,]^{\triangleleft f}$;*
6. *if $C_1 \Rightarrow C_1'$, then for any $f$: $[\,C_1\,]^{\triangleright f} \Rightarrow [\,C_1'\,]^{\triangleright f}$ and $[\,C_1\,]^{\triangleleft f} \Rightarrow [\,C_1'\,]^{\triangleleft f}$.*

The full proof is reported in Appendix A.

## 4. Behavioural Theory for AbC

In this section, we define a behavioural theory for *AbC*. We start by introducing a reduction barbed congruence, then we present an equivalent definition of a labeled bisimulation and provide a number of equational laws for it. We also show how bisimulation can be used to prove relationships between systems at different level of abstractions.

*4.1. Reduction barbed congruence*

In the behavioural theory, two terms are considered as equivalent if they cannot be distinguished by any external observer. The choice of observables is important to assess models of concurrent systems and their equivalences. For instance, in the $\pi$-calculus both message transmission and reception are considered to be observable. However, this is not the case in *AbC* because message transmission is non-blocking and thus we cannot externally observe the actual reception of a message. It is important to notice that the transition $C \xrightarrow{\Gamma \triangleright \Pi(\tilde{v})} C'$ does not necessarily mean that $C$ has performed an input action but rather it means that $C$ *might* have performed it.

Indeed, this transition might happen due to the application of one of two different rules in Table 5, namely COMP which guarantees reception and FCOMP which models non-reception. Hence, input actions cannot be observed by an external observer and only output actions are observable in *AbC*.

The minimal piece of information we can consider as *observable* from an *AbC* component is the predicate attached to the sent message. We will use the term "barb" as synonymous with observable, following the works in [10, 11]. From now onwards, we will assume that predicate $\Pi$ denotes its meaning, not its syntax. In other words, we consider predicates up to semantic equivalence $\simeq$.

**Definition 4.1** (Barb). *Let $C\downarrow_\Pi$ mean that component $C$ can send a message with some exposed environment $\Gamma$ and a predicate $\Pi'$ where $\Pi' \simeq \Pi$ and $\Pi' \not\simeq \mathsf{ff}$ (i.e., $C \xrightarrow{\Gamma \triangleright \overline{\Pi'}(\tilde{v})}$). We write $C \Downarrow_\Pi$ if $C \Rightarrow C' \downarrow_\Pi$ for some $C'$.*

To define our *reduction barbed congruence* we need to define the kind of *context* where an component $C$ can operate.

**Definition 4.2** (External context). *An external context $\mathcal{C}[\bullet]$ is a component term with a hole, denoted by $[\bullet]$. The external contexts of the AbC calculus are generated by the following grammar:*

$$\mathcal{C}[\bullet] ::= \quad [\bullet] \quad | \quad [\bullet] \| C \quad | \quad C \| [\bullet] \quad | \quad [\,[\bullet]\,]^{\triangleleft f} \quad | \quad [\,[\bullet]\,]^{\triangleright f}$$

We define notions of strong and weak barbed congruence to reason about *AbC* components following the definition of maximal sound theory by Honda and Yoshida [12]. This definition is a slight variant of Milner and Sangiorgi's barbed congruence [11] and it is also known as open barbed bisimilarity [4]. To define *reduction barbed congruence* we have to limit our attention to relations that preserve *observation* and that are preserved in any *context* and after any *reduction*.

**Definition 4.3** (Closures). *Let $\mathcal{R}$ be a binary relation over AbC-components:*

**Barb Preservation** $\mathcal{R}$ *is barb-preserving iff for every $(C_1, C_2) \in \mathcal{R}$, $C_1 \downarrow_\Pi$ implies $C_2 \Downarrow_\Pi$*

**Reduction Closure** $\mathcal{R}$ *is reduction-closed iff for every $(C_1, C_2) \in \mathcal{R}$ and predicate $\Pi$, $C_1 \xhookrightarrow{\Pi} C_1'$ implies $C_2 \xhookrightarrow{\Pi}_\tau C_2'$ for some $C_2'$ such that $(C_1', C_2') \in \mathcal{R}$*

**Context Closure** $\mathcal{R}$ *is context-closed iff for every $(C_1, C_2) \in \mathcal{R}$ and for all contexts $\mathcal{C}[\bullet]$, $(\mathcal{C}[C_1], \mathcal{C}[C_2]) \in \mathcal{R}$*

Note that, while in [12, 11, 4] *reduction closure* only takes into account *invisible actions* $\tau$, here we consider a more restrictive point of view. In fact we have a *family* of reductions, one for each kind of interaction driven by a predicate $\Pi$. This because any interaction performed over a predicate $\Pi$ is somehow *hidden* for any component that does not satisfy $\Pi$. Now, everything is in place to define reduction barbed congruence.

**Definition 4.4** (Weak Reduction Barbed Congruence). *A weak reduction barbed congruence is a symmetric relation $\mathcal{R}$ over the set of AbC-components which is barb-preserving, reduction closed, and context-closed.*

Two components are weak barbed congruent, written $C_1 \cong C_2$, if $(C_1, C_2) \in \mathcal{R}$ for some weak reduction barbed congruence relation $\mathcal{R}$. The strong reduction congruence "$\simeq$" is obtained in a similar way by replacing $\Downarrow$ with $\downarrow$ and $\hookrightarrow_\tau$ with $\hookrightarrow$.

**Lemma 4.1.** *If $C_1 \cong C_2$ then*

- $C_1 \Rightarrow C_1'$ *implies* $C_2 \Rightarrow C_{\cong C_1'}$ *where $C_{\cong C_1'}$ denotes a component that is weakly barbed congruent to $C_1'$*

- $C_1 \Downarrow_\Pi$ *iff* $C_2 \Downarrow_{\Pi'}$.

*Proof.* (We prove each statement separately)

- The proof of first item proceeds by induction on $w$ by showing that if $C_1 \Rightarrow^w C_1'$ then $C_2 \Rightarrow C_{\cong C_1'}$, where $w$ is the number of $\tau$ steps needed to move from $C_1$ to $C_1'$.
    - Base case, $w = 0$: For any $C_1$ we have that $C_1 \Rightarrow^0 C_1$. We also have that $C_2 \Rightarrow C_2$. The statement follows directly from the fact that $C_1 \cong C_2$.
    - Inductive Hypothesis: We assume that $\forall k \leq w$ if $C_1 \Rightarrow^k C_1'$ then $C_2 \Rightarrow C_{\cong C_1'}$.
    - Inductive Step: Let $C_1 \Rightarrow^{w+1} C_1'$. By definition of $\Rightarrow^{w+1}$ we have that there exists $C_1''$ such that $C_1 \xrightarrow{\tau} C_1''$ and $C_1'' \Rightarrow^w C_1'$. Since $C_1 \cong C_2$, and $\cong$ is *reduction closed* (see Definition 4.3 and Definition 4.4) we have that there exists $C_2''$ such that $C_2 \Rightarrow C_2''$ and $C_1'' \cong C_2''$. Moreover, by inductive hypothesis we have that $C_2'' \Rightarrow C_{\cong C_1'}$. Hence, $C_2 \Rightarrow C_{\cong C_1'}$ as required.

- The proof of second item follows by observing that $C_1 \Downarrow_\Pi$ if and only if $C_1 \Rightarrow C_1' \downarrow_\Pi$. Directly from the previous point we have that there exists $C_2'$ such that $C_2 \Rightarrow C_2'$ and $C_1' \cong C_2'$. Hence, since $\cong$ is *barb preserving* (see Definition 4.3 and Definition 4.4), we have that $C_2' \Downarrow_\Pi$ and, in turn, $C_2 \Downarrow_\Pi$.

□

### 4.2. Bisimulation Proof Methods

In this section, we first define a notion of labelled bisimilarity of *AbC* components, then we prove that it coincides with the reduction barbed congruence, introduced in the previous section. This "alternative" characterisation is useful to prove actual properties of *AbC* systems. In fact, barbed congruence could hardly serve the scope, since it requires testing components in every possible context.

First we need to introduce a notion of *semantic equivalence* among transition labels. The reason is that in standard process algebras, labelled bisimilarities can be defined in terms of a syntactic equivalence among labels while in *AbC* different labels may have the same meaning and impact on the context. One can consider the following two output labels $\Gamma \triangleright \overline{(a \neq 10)}(\tilde{v})$ and $\Gamma \triangleright \overline{\neg(a = 10)}(\tilde{v})$. Even if these two labels are *different*, their impact on the *context* is the same. Since this equivalence is based on the *semantic equivalence* among predicates of Definition 2.1, we use the same symbol $\simeq$ to denote this relation.

**Definition 4.5.** *We let $\simeq \subseteq SLAB \times SLAB$ be the smallest relation such that for any $\Gamma$, $\Gamma'$, $\tilde{v}$, $\tilde{w}$:*

- $\Gamma \triangleright \overline{\Pi_1}(\tilde{v}) \simeq \Gamma \triangleright \overline{\Pi_2}(\tilde{v})$, *for any $\Pi_1 \simeq \Pi_2$;*
- $\Gamma \triangleright \overline{\Pi_1}(\tilde{v}) \simeq \Gamma' \triangleright \overline{\Pi_2}(\tilde{w})$, *for any $\Pi_1 \simeq \Pi_2 \simeq \mathsf{ff}$;*
- $\Gamma \triangleright \Pi_1(\tilde{v}) \simeq \Gamma \triangleright \Pi_2(\tilde{v})$, *for any $\Pi_1 \simeq \Pi_2$.*

**Definition 4.6** (Weak Bisimulation). *A symmetric binary relation $\mathcal{R}$ over the set of AbC-components is a weak bisimulation if and only if for any $(C_1, C_2) \in \mathcal{R}$ and for any $\lambda_1$*

$$C_1 \xrightarrow{\lambda_1} C_1' \text{ implies } \exists \lambda_2 : \lambda_1 \simeq \lambda_2 \text{ such that } C_2 \xRightarrow{\widehat{\lambda_2}} C_2' \text{ and } (C_1', C_2') \in \mathcal{R}$$

*Two components $C_1$ and $C_2$ are weakly bisimilar, written $C_1 \approx C_2$ if there exists a weak bisimulation $\mathcal{R}$ relating them.*

It is worth noting that *strong bisimulation* and *strong bisimilarity* ($\sim$) can be defined similarly, only $\stackrel{\widehat{\lambda_2}}{\Longrightarrow}$ is replaced by $\stackrel{\lambda_2}{\longrightarrow}$. It is easy to prove that $\sim$ and $\approx$ are equivalence relations by relying on the classical arguments of [13]. However, our bisimilarities enjoy a much more interesting property: closure under any external context.

The following lemmas state that our weak labelled bisimilarities is preserved by parallel composition, and restriction. Similar lemmas do hold also for the strong variant. The proofs of these lemmas are reported in Appendix A.

**Lemma 4.2** ($\approx$ is preserved by parallel composition). *If $C_1$ and $C_2$ are two components, we have that $C_1 \approx C_2$ implies $C_1 \| C \approx C_2 \| C$ for all components $C$.*

**Lemma 4.3** ($\approx$ is preserved by restriction). *If $C_1$ and $C_2$ are two components, we have that $C_1 \approx C_2$ implies $[\, C_1 \,]^{\lhd f} \approx [\, C_2 \,]^{\lhd f}$ and $[\, C_1 \,]^{\rhd f} \approx [\, C_2 \,]^{\rhd f}$ for any $f$.*

As an immediate consequence of Lemma 4.2 and Lemma 4.3, we have that $\approx$ is a congruence relation (i.e., closed under any external $AbC$ context). Notably, similar lemmas do hold also for $\sim$.

We are now ready to show how weak bisimilarity can be used as a proof technique for reduction barbed congruence.

**Theorem 4.1** (Soundness). *$C_1 \approx C_2$ implies $C_1 \cong C_2$, for any two components $C_1$ and $C_2$.*

*Proof.* It is sufficient to prove that bisimilarity is barb-preserving, reduction-closed, and context-closed.

- (Barb-preservation): By the definition of the barb $C_1 \downarrow_\Pi$ if $C_1 \xrightarrow{\Gamma \rhd \overline{\Pi}(\tilde{v})}$ for an output label $\Gamma \rhd \overline{\Pi}(\tilde{v})$ with $\Pi \neq \mathrm{ff}$. As $(C_1 \approx C_2)$, we have that also $C_2 \stackrel{\Gamma \rhd \overline{\Pi}(\tilde{v})}{\Longrightarrow}$ and $C_2 \Downarrow_\Pi$.

- (Reduction-closure): Let $C_1 \stackrel{\Pi}{\hookrightarrow} C_1'$. This means that there exist $\Gamma$, $\tilde{v}$ and $\Pi'$ such that $C_1 \xrightarrow{\Gamma \rhd \overline{\Pi'}(\tilde{v})} C_1'$ and $\Pi \simeq \Pi'$. As $(C_1 \approx C_2)$, then there exists $C_2'$ such that $C_2 \stackrel{\widehat{\Gamma \rhd \Pi''(\tilde{v})}}{\Longrightarrow} C_2'$ with $\Pi' \simeq \Pi''$ and $(C_1' \approx C_2')$. Hence, $C_2 \stackrel{\Pi}{\hookrightarrow} C_2'$ and $(C_1' \approx C_2')$.

- (Context-closure): Let $(C_1 \approx C_2)$ and let $\mathcal{C}[\bullet]$ be an arbitrary $AbC$-context. By induction on the structure of $\mathcal{C}[\bullet]$ and using Lemma 4.2 and Lemma 4.3, we have that $\mathcal{C}[C_1] \approx \mathcal{C}[C_2]$.

In conclusion, we have that $C_1 \cong C_2$ as required. $\square$

This soundness theorem allow us to use bisimilarity when we have to prove that two $AbC$ components are *barbed equivalent*. We want now to study *completeness* in order to show that bisimilarity is more than a proof technique, it rather represents an alternative characterisation of reduction barbed congruence.

**Lemma 4.4** (Completeness). *$C_1 \cong C_2$ implies $C_1 \approx C_2$, for any two components $C_1$ and $C_2$.*

*Proof.* It is sufficient to prove that the relation $\mathcal{R} = \{(C_1, C_2) \mid \text{ such that } (C_1 \cong C_2)\}$ is a bisimulation. For this reason we have to show that:

- $\mathcal{R}$ is symmetric;

- for each $(C_1, C_2) \in \mathcal{R}$ and for any $\lambda_1$

$$C_1 \xrightarrow{\lambda_1} C_1' \text{ implies } \exists \lambda_2 : \lambda_1 \simeq \lambda_2 \text{ such that } C_2 \stackrel{\widehat{\lambda_2}}{\Longrightarrow} C_2' \text{ and } (C_1', C_2') \in \mathcal{R}$$

The first item derives directly from the fact that $\cong$ is symmetric. To prove the second item we have to consider three cases:

*Case 1:* $\lambda_1 = \Gamma \triangleright \overline{\Pi}(\tilde{v})$ and $\Pi \simeq \mathsf{ff}$. From Definition 4.4, we have that $C_1 \cong C_2$ is *reduction preserving*. This means that if $C_1 \overset{\mathsf{ff}}{\hookrightarrow} C_1'$ then there exists $C_2'$ such that $C_2 \overset{\mathsf{ff}}{\hookrightarrow}_\tau C_2'$, that is there exists $\lambda_2 = \Gamma' \triangleright \overline{\Pi''}(\tilde{v}') \simeq \lambda_1$ such that $C_2 \overset{\widehat{\lambda_2}}{\Rightarrow} C_2'$ and $C_1' \cong C_2'$.

*Case 2:* $\lambda_1 = \Gamma \triangleright \overline{\Pi}(\tilde{v})$ and $\Pi \not\simeq \mathsf{ff}$. We have to prove that if $C_1 \overset{\lambda_1}{\longrightarrow} C_1'$ there exists $\lambda_2 \simeq \lambda_1$ such that $C_2 \overset{\lambda_2}{\Rightarrow} C_2'$ and $C_1' \cong C_2'$.

Let us consider the following context $\mathcal{C}[\bullet]$:

$$\mathcal{C}[\bullet] = \bullet \parallel \Gamma_\Pi : (\Pi_\Gamma \wedge (\tilde{x} = \tilde{v}))(\tilde{x}).((\mathsf{tt})@(\mathsf{b} = \mathsf{tt}).0 + (\mathsf{tt})@\mathsf{ff}.0)$$

where $\Gamma_\Pi \in \mathcal{M}(\Pi)$, $\Pi_\Gamma$ is a predicate uniquely identifying $\Gamma$, i.e. $\mathcal{M}(\Pi_\Gamma) = \{\Gamma\}$, while $\mathsf{b}$ is an attribute occurring neither in $C_1$ nor in $C_2$. Let $A$ be the (finite) set of attributes occurring in $C_1$ and $C_2$, predicate $\Pi_\Gamma$ can be defined as:

$$\Pi_\Pi = \bigwedge_{\mathsf{a} \in A} (\mathsf{a} = \Gamma(\mathsf{a}))$$

Since $\cong$ is *context closed*, we have that $\mathcal{C}[C_1] \cong \mathcal{C}[C_2]$. Moreover, by applying rules of Table 5, we have that:

$$\mathcal{C}[C_1] \overset{\lambda_1}{\longrightarrow} C_1' \parallel \Gamma_\Pi : (\mathsf{tt})@(\mathsf{b} = \mathsf{tt}).0 + (\mathsf{tt})@\mathsf{ff}.0$$

This implies that

$$\mathcal{C}[C_1] \overset{\Pi}{\hookrightarrow} C_1' \parallel \Gamma_\Pi : (\mathsf{tt})@(\mathsf{b} = \mathsf{tt}).0 + (\mathsf{tt})@\mathsf{ff}.0$$

Since $\cong$ is reduction closed we have that

$$\mathcal{C}[C_2] \overset{\Pi}{\hookrightarrow}_\tau C' \quad \text{and} \quad C_1' \parallel \Gamma_\Pi : (\mathsf{tt})@(\mathsf{b} = \mathsf{tt}).0 + (\mathsf{tt})@\mathsf{ff}.0 \cong C'$$

Since $\mathsf{b}$ does not occur in $C_2$, and $C' \Downarrow_{(\mathsf{b}=\mathsf{tt})}$, we can infer that there exists $\lambda_2 = \Gamma' \triangleright \overline{\Pi'}(\tilde{v}')$ such that $\Pi \simeq \Pi'$:

$$\mathcal{C}[C_2] \overset{\lambda_2}{\Rightarrow} C_2' \parallel \Gamma_\Pi : (\mathsf{tt})@(\mathsf{b} = \mathsf{tt}).0 + (\mathsf{tt})@\mathsf{ff}.0$$

where $\Gamma' \models \Pi_\Gamma$ and $\tilde{v}' = \tilde{v}$. This implies that $\Gamma' \in \Pi_\Gamma = \{\Gamma\}$ and $\lambda_1 \simeq \lambda_2$.

We have now observe that:

$$C_1' \parallel \Gamma_\Pi : (\mathsf{tt})@(\mathsf{b} = \mathsf{tt}).0 + (\mathsf{tt})@\mathsf{ff}.0 \overset{\mathsf{ff}}{\hookrightarrow} C_1' \parallel \Gamma_\Pi : 0$$

We can use again the fact that $\cong$ is *reduction closed* to have that:

$$C_2' \parallel \Gamma_\Pi : (\mathsf{tt})@(\mathsf{b} = \mathsf{tt}).0 + (\mathsf{tt})@\mathsf{ff}.0 \overset{\mathsf{ff}}{\hookrightarrow}_\tau C_2'' \parallel \Gamma_\Pi : 0$$

where $C_2' \Rightarrow C_2''$ and $C_1' \parallel \Gamma_\Pi : 0 \cong C_2'' \parallel \Gamma_\Pi : 0$. We can observe that $C_1' \parallel \Gamma_\Pi : 0 \cong C_2'' \parallel \Gamma_\Pi : 0$ if and only if $C_1' \cong C_2''$.

Finally, we have that $C_2 \overset{\lambda_2}{\Rightarrow} C_2''$ and $C_1' \cong C_2''$.

*Case 3:* $\lambda_1 = \Gamma \triangleright \Pi(\tilde{v})$. In this case the proof proceeds like for the previous one by considering the following context $\mathcal{C}[\bullet]$:

$$\mathcal{C}[\bullet] = \bullet \parallel \Gamma : (\Pi)@\tilde{v}.((\mathtt{tt})@(\mathtt{b} = \mathtt{tt}).0 + (\mathtt{tt})@\mathtt{ff}.0)$$

$\square$

**Theorem 4.2** (Characterisation). *Bisimilarity and reduction barbed congruence coincide.*

*Proof.* As a direct consequence of Theorem 4.1 and Lemma 4.4, we have that weak bisimilarity and weak reduction barbed congruence coincide. $\square$

The proof for the strong variant of equivalence (i.e., $C_1 \simeq C_2$ coincides with $C_1 \sim C_2$) follows in a similar way and it is omitted for the sake of brevity.

## 5. Bisimulations at work

In the previous section we proved that bisimilarity is a congruence relation for all external *AbC* contexts, i.e., system level contexts as described in Definition 4.2. In this section we show that, due to the dependencies of processes on the attribute environment, almost all process-level operators do not preserve bisimilarity, the only exception being the awareness operator. However, this problem can be solved by closing bisimilarity under any possible substitution as we will see later. Notice that our bisimilarity is still a congruence because it is defined at the level of system components and thus only external contexts matter. The rest of the section concentrates on other properties and equational laws exhibited by bisimilarity. Unless stated otherwise, the properties hold for both strong and weak bisimilarity.

*5.1. Equational Laws for* AbC *Bisimulation*

As mentioned above, weak bisimilarity is not preserved by most process level operators.

**Remark 5.1.** *For some attribute environment $\Gamma$, an interface $I$, and two processes $P$ and $Q$ where $\Gamma :_I P \approx \Gamma :_I Q$, we have that*

1. $\Gamma :_I \alpha.P \not\approx \Gamma :_I \alpha.Q$ *for some action $\alpha$*
2. $\Gamma :_I P|R \not\approx \Gamma :_I Q|R$ *for some process $R$*
3. $\Gamma :_I \langle \Pi \rangle P \approx \Gamma :_I \langle \Pi \rangle Q$ *for every predicate $\Pi$*
4. $\Gamma :_I \alpha.[a := E]P \not\approx \Gamma :_I \alpha.[a := E]Q$ *for some update $[a := E]$*

*Proof.* Let $C_1 = \Gamma :_I \overbrace{\langle \mathtt{this}.a = w \rangle (v')@\Pi.0}^{P}$ where $\Gamma(a) = v$, $C_2 = \Gamma :_I \overbrace{0}^{Q}$, and $R = ()@\mathtt{ff}.[a := w]0$. It is easy to see that $C_1 \approx C_2$, because both components are not able to progress. Notice that $(\mathtt{this}.a = w) \not\models \Gamma$.

1. The statement, $\Gamma :_I \alpha.P \not\approx \Gamma :_I \alpha.Q$ for some action $\alpha$, is a direct consequence of the first statement. For instance, consider an input prefix of the following form $(\mathtt{tt})(w)$.
2. The statement, $\Gamma :_I P|R \not\approx \Gamma :_I Q|R$ for some process $R$, holds easily from our example when we put the process $R$ in parallel of the processes $P$ and $Q$.
3. The statement, $\Gamma :_I \langle \Pi \rangle P \approx \Gamma :_I \langle \Pi \rangle Q$ for every predicate $\Pi$, is a direct sequence of operational rules for the awareness operator.
4. The last statement holds easily with the following update $[a := w]$.

$\square$

It should be noted that if we close bisimilarity under substitutions by definition, all of the statements in Remark 5.1 can be proved to preserve bisimilarity. The definition would be a slight variant of the notion of full bisimilarity proposed by Sangiorgi and Walker in [4]. In this way, the components $C_1$ and $C_2$ in the proof above are no longer bisimilar since they are not equivalent after substitution $[v/w]$. However, the new notion of bisimilarity induced by the closure is finer than the one proposed in this article.

The following remark shows that, as expected, non-deterministic choice does not preserve bisimilarity. The reason is related to the fact that input transitions cannot be observed. Below we explain the issue with a concrete example.

**Remark 5.2.** *For some attribute environment $\Gamma$, an interface $I$, and two processes $P$ and $Q$ where $\Gamma:_I P \approx \Gamma:_I Q$, we have that $\Gamma:_I P + R \not\approx \Gamma:_I Q + R$ for any process $R$*

*Proof.* Let $C_1 = \Gamma:_I \Pi_1(x).0$, $C_2 = \Gamma:_I \Pi_2(x).0$, and $R = (v)@\Pi.0$. Though the receiving predicates for both components are different we still have that $C_1 \approx C_2$ and this is because that input actions are not perceived. When a message $\Gamma' \triangleright \overline{\Pi_3}(w)$ arrives, where $\Gamma \downarrow I \models \Pi_3$, $\Gamma' \models [\![\Pi_1[w/x]]\!]_\Gamma$ and $\Gamma' \not\models [\![\Pi_2[w/x]]\!]_\Gamma$, component $C_1$ applies rule COMP and evolves to $\Gamma:_I 0$ while component $C_2$ applies rule FCOMP and stays unchanged. Both transitions carry the same label and again $\Gamma:_I 0$ and $\Gamma:_I \Pi_2(x).0$ are equivalent for a similar reason. An external observer cannot distinguish them.

Now if we allow mixed choice within a single component, then one can distinguish between $\Pi_1(x)$ and $\Pi_2(x)$.

$$\Gamma:_I \Pi_1(x).0 + R \not\approx \Gamma:_I \Pi_2(x).0 + R$$

Assume that the message $\Gamma' \triangleright \overline{\Pi_3}(w)$ is arrived, we have that:

$$\Gamma:_I \Pi_1(x).0 + R \xrightarrow{\Gamma' \triangleright \Pi_3(w)} \Gamma:_I 0 \xcancel{\xrightarrow{\Gamma \downarrow I \triangleright \overline{\Pi}(v)}}$$

while

$$\Gamma:_I \Pi_2(x).0 + R \xrightarrow{\Gamma' \triangleright \Pi_3(w)} \Gamma:_I \Pi_2(x).0 + R \xrightarrow{\Gamma \downarrow I \triangleright \overline{\Pi}(v)} \Gamma:_I 0$$

However, this is obvious since our relation is defined at the system-level. So it abstracts from internal behaviour and characterises the behaviour of $AbC$ systems from an external observer point of view. In practice this is not a problem since mixed choice (i.e., nondeterministic choice between input and output actions) is very hard to be implemented. □

The following lemmas prove useful properties about $AbC$ operators (i.e., parallel composition is commutative, associative, ...). We omit their proofs; they follows directly from the operational semantics of $AbC$ that we presented in Section 3.

**Lemma 5.1** (Parallel composition).

- $C_1 \| C_2 \approx C_2 \| C_1$
- $(C_1 \| C_2) \| C_3 \approx C_1 \| (C_2 \| C_3)$
- $\Gamma:_I 0 \| C \approx C$

**Lemma 5.2** (Non-deterministic choice).

- $\Gamma:_I P_1 + P_2 \approx \Gamma:_I P_2 + P_1$
- $\Gamma:_I (P_1 + P_2) + P_3 \approx \Gamma:_I P_1 + (P_2 + P_3)$
- $\Gamma:_I P + 0 \approx \Gamma:_I P$
- $\Gamma:_I P + P \approx \Gamma:_I P$
- $\Gamma:_I \langle\Pi\rangle(P + Q) \approx \Gamma:_I \langle\Pi\rangle P + \langle\Pi\rangle Q$

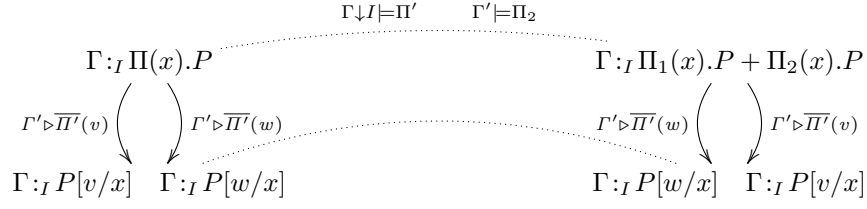

Figure 1: The relationship between the "or" predicate and the non-deterministic choice

**Lemma 5.3** (Interleaving).

- $\Gamma:_I P_1|P_2 \approx \Gamma:_I P_2|P_1$
- $\Gamma:_I (P_1|P_2)|P_3 \approx \Gamma:_I P_1|(P_2|P_3)$
- $\Gamma:_I P|0 \approx \Gamma:_I P$

**Lemma 5.4** (Awareness).

- $\Gamma:_I \langle \mathsf{ff} \rangle P \approx \Gamma:_I 0$
- $\Gamma:_I \langle \mathsf{tt} \rangle P \approx \Gamma:_I P$
- $\Gamma:_I \langle \Pi_1 \rangle \langle \Pi_2 \rangle P \approx \Gamma:_I \langle \Pi_1 \wedge \Pi_2 \rangle P$

**Lemma 5.5** (Silent components cannot be observed). *Let $Act(P)$ denote the set of actions in process $P$. If $Act(P)$ does not contain any output action, then:*

$$\Gamma:_I P \approx \Gamma:_I 0$$

*Proof.* The proof follows from the fact that components with no external side-effects (i.e., do not exhibit barbs) cannot be observed. When $Act(P)$ does not contain output actions, component $\Gamma:_I P$ can either make silent moves, which component $\Gamma:_I 0$ can mimic by simply doing nothing, or input a message, which component $\Gamma:_I 0$ can mimic by discarding the message. □

*5.2. Proving equivalence of AbC systems*

Now we proceed with a few examples to provide evidence of interesting features of the *AbC* calculus.

**Example 5.1.** *We have that $C_1 \approx C_2$ when $C_1 = \Gamma:_I \Pi(x).P$, $C_2 = \Gamma:_I \Pi_1(x).P + \Pi_2(x).P$ and $\Pi \simeq \Pi_1 \vee \Pi_2$.*

Clearly, components $C_1$ and $C_2$ are bisimilar because any message, accepted by $C_2$, can also be accepted by $C_1$ and vice versa. After a successful input both components proceed with the same continuation process $P[v/x]$. For instance, consider the message $\Gamma' \triangleright \overline{\Pi_1}(v)$ in which $\Gamma'$ is only satisfied by predicate $\Pi_2$, it is still satisfied by predicate $\Pi$. The overlapping between the input and the non-deterministic choice constructs is clear in this scenario. For this special case we can replace the non-deterministic choice with an "or" predicate while preserving the observable behaviour. The intuition is illustrated in Figure 1.

It is worth noting that as a corollary of the above equivalence we have:

$$\Gamma:_I \Pi_1(\tilde{x}).P + \cdots + \Pi_n(\tilde{x}).P \approx \Gamma:_I (\Pi_1 \vee \Pi_2 \vee \cdots \vee \Pi_n)(\tilde{x}).P$$

**Example 5.2.** $\Gamma_1:_I (E_1)@\Pi.P \approx \Gamma_2:_I (E_2)@\Pi.P$ *if and only if* $\Gamma_1 \downarrow I = \Gamma_2 \downarrow I$ *and* $[\![E_1]\!]_{\Gamma_2} = [\![E_2]\!]_{\Gamma_1}$.

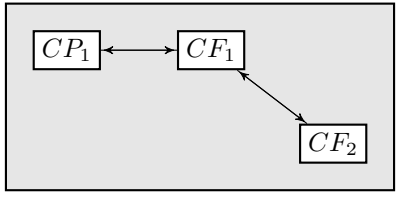

Figure 2: The system with assumptions about the network topology

It is clear that even if $\Gamma_1 \neq \Gamma_2$, these components are still bisimilar since their interfaces and exposed messages are equivalent. This is an important property which ensures that components need not to have isomorphic attribute environments to be equivalent. The intuition is that components can control what attribute values to be exposed to the communication partners. In some sense the component has the power of selecting the criteria in which its communicated messages can be filtered.

Now we show some interesting properties about the restriction mechanism in $AbC$. The restriction mechanism in $AbC$ where a predicate can be partially exposed is very useful in describing collective behaviour from a global point of view. This means that local interactions are hidden from an external observer which can only observe system level behaviour (collective-behaviour). In the next example we show the expressive power of name restriction in a more elaborated scenario.

**Example 5.3.** *We consider two types of forwarding components, a source component $CP = \Gamma_p :_I P$ and an intermediate component $CF = \Gamma_i :_I F$ where the behaviour of processes $P$ and $F$ is defined below.*

$$P \triangleq (\texttt{this}.\text{id},\ \tilde{v})@(\Pi_1 \vee (\text{role} = fwd)).0$$

$$F \triangleq (x \in \texttt{this}.\text{nbr})(x,\ \tilde{y}).(x,\ \tilde{y})@(\text{role} = fwd).(x,\ \tilde{y})@\Pi_1.0$$

Process $P$ sends an advertisement message to all components that either satisfy predicate $\Pi_1$ where $\Pi_1 = (\text{role} = client)$ or have a forwarder role (i.e., $(\text{role} = fwd)$). Process $F$ may receive an advertisement from a neighbour, after that it first sends it to nearby forwarders and then sends the advertisement to nearby clients. The scenario is simplified to allow at most two hops from the source component.

The goal of the source component is to ensure that its advertisement message reaches all clients across the network. To prove if the above specification guarantees this property[2], we first need to fix the topology of the network as reported in Figure 2. For the sake of simplicity we will only consider a network of one source $CP_1 = \Gamma_p :_{\{role\}} P$ and two forwarders $CF_1 = \Gamma_1 :_{\{role\}} F$ and $CF_2 = \Gamma_2 :_{\{role\}} F$. Notice that the interface of these components contains the role attribute. We assume short-range communication where $CP_1$ messages can reach to $CF_1$ and $CF_2$ can only receive the messages when $CF_1$ forwards them. Assume that initially the attribute environments $\Gamma_p$, $\Gamma_1$ and $\Gamma_2$ are defined as follows:

$\Gamma_p = \{(\text{role}, fwd),\ (\text{id}, p),\ (\text{nbr}, \{f_1, f_4\})\}, \quad \Gamma_1 = \{(\text{id}, f_1),\ (\text{role}, fwd),\ (\text{nbr}, \{p, f_2, f_3\})\}$
$\Gamma_2 = \{(\text{id}, f_2),\ (\text{role}, fwd),\ (\text{nbr}, \{f_1\})\}$

To avoid interference with other components running in parallel we can use *restriction operators* to guarantee that interactions between the source and the forwarders are *private*. The full system is represented by the component $N$ as defined below:

$$N = [\ CP_1 \parallel CF_1 \parallel CF_2\ ]^{\triangleright f_{\mathsf{fwd}}}$$

where $f_{\mathsf{fwd}}(\Gamma, \tilde{v})$ yields $(\text{role} \neq \mathsf{fwd})$ and guarantees that only non-forwarders can receive a message outside the boundaries of the restriction.

---
[2]The results in this scenario only hold for weak bisimulation.

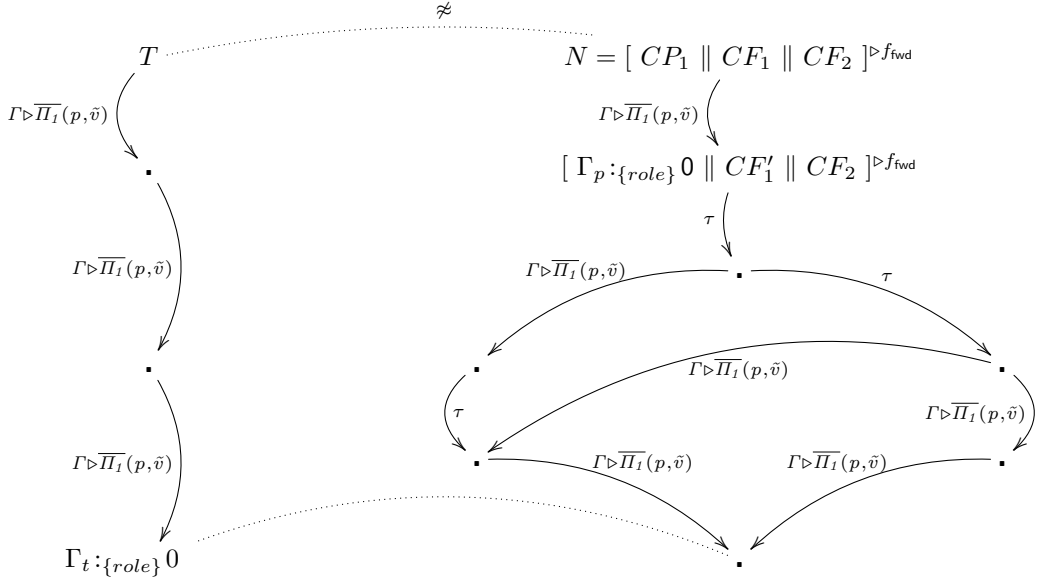

Figure 3: System $N$ simulates the test component $T$, but initial interference is possible, Hence $N \not\approx T$

The behaviour of $N$ without any interventions from other source components is reported on the right side of Figure 3. The source component $CP_1$ initiates the interaction by sending an advertisement to nearby clients and forwarders and evolves to $\Gamma_p :_{\{role\}} 0$. Forwarder $CF_1$ receives the message and evolves to $CF_1'$. The overall system $N$ applies rule RESO and evolves to $[\ \Gamma_p :_{\{role\}} 0 \parallel CF_1' \parallel CF_2\ ]^{\triangleright f_{fwd}}$ with the label $\Gamma \triangleright \overline{(\Pi_1 \vee (\text{role} = fwd)) \wedge ((\text{role} \neq fwd))}(p, \tilde{v})$ which is equivalent to $\Gamma \triangleright \overline{\Pi_1}(p, \tilde{v})$. Notice that $\Gamma$ is equivalent to $\Gamma_p \downarrow I$. The forwarder $CF_1'$ forwards its message to nearby forwarders, in our case this is $CF_2$. The overall system applies again rule RESO and evolves to $[\ \Gamma_p :_{\{role\}} 0 \parallel CF_1'' \parallel CF_2'\ ]^{\triangleright f_{fwd}}$ with the label $\Gamma \triangleright \overline{(\text{role} = fwd) \wedge (\text{role} \neq fwd)}(p, \tilde{v})$. This message is private and is perceived externally as a silent move. The overall system terminates after another internal action (performed by the second forwarder) and by emitting the message, $\Gamma \triangleright \overline{\Pi_1}(p, \tilde{v})$[3], two more times, one from each forwarder. By applying the rule RESO twice, the system evolves to $[\ \Gamma_p :_{\{role\}} 0 \parallel \Gamma_1 :_{\{role\}} 0 \parallel \Gamma_2 :_{\{role\}} 0\ ]^{\triangleright f_{fwd}}$.

To prove that the advertising message is propagated to all clients in the network it is sufficient to show that each forwarder takes its turn in spreading the message. Formally it is sufficient to prove that the behaviour of the overall system is bisimilar to the behaviour of a test component $T$ with $\Gamma_t(\text{role}) = fwd$, defined below, which is able to send the same message three times sequentially and then terminates.

$$T = \Gamma_t :_{\{role\}} (p,\ \tilde{v})@\Pi_1.(p,\ \tilde{v})@\Pi_1.(p,\ \tilde{v})@\Pi_1.0$$

Figure 3 shows that system $N$ weakly simulates component $T$, but they are not bisimilar, i.e., $T \not\approx N$. This is because a forwarder is initially prepared to accept any message $(a, \tilde{v})$ such that $a$ coincides with the id of one of its neighbours. For instance if we put another component, say $CP_2 = \Gamma_h :_{\{\}} (f_3,\ \tilde{w})@(\text{tt}).0$, there is a possibility that $CF_1$ first receives a message from $CP_2$ and the system can evolve as follows:

$$N \| CP_2 \xrightarrow{\{\} \triangleright \overline{\text{tt}}(f_3, \tilde{w})} \xrightarrow{\Gamma \triangleright \overline{\Pi_1}(p, \tilde{v})} \xrightarrow{\Gamma \triangleright \overline{\text{ff}}()} \xrightarrow{\Gamma \triangleright \overline{\text{ff}}()} \xrightarrow{\Gamma \triangleright \overline{\Pi_1}(f_3, \tilde{w})} \xrightarrow{\Gamma \triangleright \overline{\Pi_1}(f_3, \tilde{w})}$$

---

[3] Note that since role can only assume the values $client$ and $fwd$, we have that $(\text{role} = client) \wedge (\text{role} \neq fwd) \stackrel{\frown}{=} \text{role} = client) = \Pi_1$

while

$$T \| CP_2 \xrightarrow{\{\} \triangleright \overline{\text{tt}}(f_3, \tilde{w})} \xrightarrow{\Gamma \triangleright \overline{\Pi_1}(p, \tilde{v})} \xrightarrow{\Gamma \triangleright \overline{\Pi_1}(p, \tilde{v})} \xrightarrow{\Gamma \triangleright \overline{\Pi_1}(p, \tilde{v})}$$

and it is easy to see that $N \| CP_2 \not\approx T \| CP_2$. One way to avoid interference and ensure that the property holds is to limit input capabilities of $N$:

$$N = [\ [\ CP_1 \parallel CF_1 \parallel CF_2\ ]^{\triangleright f_{\text{fwd}}}\ ]^{\triangleleft g_*}$$

where $g_*(\Gamma, \tilde{v}) = \text{ff}$ for any $\Gamma$ and $\tilde{v}$. Under this restriction $N$ is not able to receive any message and $N \approx T$.

## 6. Encoding channel-based interaction

In this section, we provide evidences of the expressive power of the $AbC$ calculus by discussing how other interaction patterns can be naturally expressed in $AbC$ and advocate the use of attribute-based communication as a unifying framework to encompass different communication models. First, we discuss how *group-based* [14, 15, 5] and *publish/subscribe-based* [16, 17] interaction patterns can be naturally rendered in $AbC$[4], then we concentrate on the encoding of a process calculus where processes interact via broadcasting channels and formally prove its correctness.

In the group-based model, when an agent wants to send a message, it attaches the group name/id in the message and only members of that group can receive the message when it is propagated. To model this interaction pattern in $AbC$, group names can be rendered as attributes and the constructs for joining or leaving a given group can be modelled as attribute updates.

In the publish/subscribe model, there are two types of agents: publishers and subscribers and there is an exchange server that mediates the interaction between them. Publishers produce messages tagged with topics and send them to the exchange server which is responsible for filtering and forwarding these messages to interested subscribers. Subscribers simply register their interests to the exchange server and based on their interests they receive messages. A natural modeling of the topic-based publish/subscribe model [17] into $AbC$ can be accomplished by allowing publishers to broadcast messages with "tt" predicates (i.e., satisfied by all subscribers) and only subscribers can check the compatibility of the exposed publishers attributes with their subscriptions.

In the next subsection we will show in full details how it is possible to model channel based communication in $AbC$, but first we would like to spend some words about the difficulties that channel based calculi have in mimicking situations that are naturally expressed in $AbC$.

In contrast to the classical process calculi, where senders and receivers have to agree on an explicit channel or name, $AbC$ relies on the satisfaction of predicates over attributes or communicated values for establishing an interaction. Attribute values in $AbC$ can be modified by means of internal actions. Changing attributes values makes it possible to have opportunistic interactions among components in the sense that an attribute update might provide new opportunities of interaction, because the selection of interaction partners depends on predicates over the attributes they expose. Changing the values of these attributes implies changing the set of possible partners and this makes it quite natural to model adaptivity in $AbC$. Offering this possibility in channel-based process calculi is not easy.

Indeed, we would like to argue that encoding the notion of interdependence between co-located processes in channel-based process calculi is very difficult if not impossible. See the following simple $AbC$ system:

$$\Gamma_1 :_{\{b\}} (msg)@(\text{tt}) \| \\ \Gamma_2 :_{I'} (()@\text{ff}.[\text{this}.a := 5]P \mid (b \leq \text{this}.a)(x).Q)$$

If we assume that initially $\Gamma_1(b) = 3$ and $\Gamma_2(a) = 2$, we have that changing the value of the local attribute $a$ to "5" by the first process in the second component gives it an opportunity of receiving the message "$msg$" from the process in the first component. One would argue that using restriction to hide local communication

---

[4]Further details about the way these two communication paradigms are modelled in $AbC$ can be found in [9].

(Component Level)

$$(|G|)_c \triangleq [(|G|)_p] \qquad (|P_1 \| P_2|)_c \triangleq (|P_1|)_c \| (|P_2|)_c$$

(Process Level)

$$(|\mathtt{nil}|)_p \triangleq 0 \qquad (|\tau.G|)_p \triangleq ()@\mathtt{ff}.(|G|)_p$$

$$(|a(\tilde{x}).G|)_p \triangleq \Pi(y, \tilde{x}).(|G|)_p$$
$$\texttt{with} \quad \Pi = (y = a) \quad \texttt{and} \quad y \notin n((|G|)_p)$$

$$(|\bar{a}\tilde{x}.G|)_p \triangleq (a, \tilde{x})@(\mathtt{tt}).(|G|)_p$$

$$(|(rec\ A\langle\tilde{x}\rangle).G\langle\tilde{y}\rangle|)_p \triangleq (A(\tilde{x}) \triangleq (|G|)_p)$$
$$\texttt{where} \quad fn((|G|)_p) \subseteq \{\tilde{x}\}$$
$$(|G_1 + G_2|)_p \triangleq (|G_1|)_p + (|G_2|)_p$$

Table 6: Encoding b$\pi$-calculus into $AbC$

and bound input/output actions would be sufficient to encode such kind of behaviors in channel-based process calculi. However, this is not the case because bound input/output actions can engage in communication only when they are instantiated with concrete channel names. In the example above, the input action of the second process of the second component is always enabled. This means that before the update, an input is available on the predicate $y \leq 2$ and after the update it is available on the predicate $y \leq 5$.

6.1. Encoding of broadcasting channels into AbC

We consider now, in some details, the issue of encoding one-to-many *channel-based interaction* in the $AbC$ calculus. It may seem tempting to model a channel name as an attribute in $AbC$, however it turns out not to be the correct approach. The reason is that in channel-based communication, a channel, where the exchange happens, is instantly enabled at the time of interaction and is disabled afterwards. This feature is not present in $AbC$ since attributes are persistent in the attribute environment and cannot be disabled at any time (i.e., attribute values can always be checked against sender's predicates). However, this is not a problem because we can exploit the fact that the receiving predicates in $AbC$ can check the values contained in the received message. The key idea is to use structured messages to select communication partners where the name of the channel is rendered as the first element of messages; receivers only accept those messages that have attached channels that match their receiving channels. Actually, attributes do not play any role in such interaction so we can consider components with empty environments and empty interfaces i.e., $\emptyset :_\emptyset P$.

To show feasibility of the approach just outlined, we encoded in $AbC$ a process algebra, named b$\pi$-calculus, (inspired by CBS [6] and [7]). We consider the set of processes $P$ as shown below.

$$P ::= G \mid P_1 \| P_2$$
$$G ::= \mathtt{nil} \mid a(\tilde{x}).G \mid \bar{a}\tilde{x}.G \mid G_1 + G_2 \mid (rec\ A\langle\tilde{x}\rangle.G)\langle\tilde{y}\rangle$$

As reported in Table 6, the encoding of a process $P$ is rendered as an $AbC$ component $(|P|)_c$ with $\Gamma = I = \emptyset$. In what follows, we use $[G]$ to denote a component with empty $\Gamma$ and $I$, i.e., $\emptyset :_\emptyset \mathsf{G}$. Notice that $(|G|)_c$ encodes a sequential process while $(|P|)_c$ encodes the parallel composition of sequential processes. The channel is rendered as the first element in the sequence of values. For instance, in the output action $(a, \tilde{x})@(\mathtt{tt})$, $a$ represents a channel name, so the input action $(y = a)(y, \tilde{x})$ will always check the first element of the received values to decide whether to accept or discard the message.

6.2. Correctness of the encoding

In this section, we provide the correctness proof of the encoding presented in Section 6.1. We start by defining the properties that we would like our encoding to preserve. Basically, when translating a term from $b\pi$-calculus into $AbC$, we would like that the translation is compositional and independent from contexts; is independent from the names of the source term (i.e., name invariance); preserves parallel composition (i.e., homomorphic w.r.t. '|'); is faithful in the sense that it preserves the observable behavior (i.e., barbs) and divergence. Moreover, the encoding has to translate output (input) actions of $b\pi$-terms into corresponding output (input) $AbC$ actions, and has to preserve the operational correspondence between the source and target calculus. This includes that the translation should be complete (i.e., every computation of the source term can be mimicked by its translation) and it should be sound (i.e., every computation of a translated term corresponds to some computation of its source term).

**Definition 6.1** (Divergence). *$P$ diverges, written $P \Uparrow$, iff $P \to^\omega$ where $\omega$ denotes an infinite number of reductions.*

**Definition 6.2** (Uniform Encoding). *An encoding $( \mathbin{.} ) : \mathcal{L}_1 \to \mathcal{L}_2$ is uniform if it enjoys the following properties:*

1. *(Homomorphic w.r.t. parallel composition):* $( P \| Q ) \triangleq ( P ) \| ( Q )$

2. *(Name invariance):* $( P\sigma ) \triangleq ( P )\sigma$, *for any permutation of names $\sigma$.*

3. *(Faithfulness):* $P \Downarrow_1$ iff $( P ) \Downarrow_2$; $P \Uparrow_1$ iff $( P ) \Uparrow_2$

4. *Operational correspondence*

    1. *(Operational completeness): if $P \to_1 P'$ then $( P ) \to^*_2 \simeq_2 ( P' )$ where $\simeq$ is the strong barbed equivalence of $\mathcal{L}_2$.*

    2. *(Operational soundness): if $( P ) \to_2 Q$ then there exists a $P'$ such that $P \to^*_1 P'$ and $Q \to^*_2 \simeq_2 ( P' )$, where $\simeq$ is the strong barbed equivalence of $\mathcal{L}_2$.*

**Lemma 6.1** (Operational Completeness). *If $P \to_{b\pi} P'$ then $(P)_c \to^* \simeq (P')_c$.*

Now we provide a sketch of the proof of the operational completeness and we report its full details in the Appendix B.

*Proof.* (Sketch) The proof proceeds by induction on the shortest transition of $\to_{b\pi}$. We have several cases depending on the structure of the term $P$. We only consider the case of parallel composition when communication happens: $P_1 \| P_2 \xrightarrow{\bar{a}\tilde{z}} P'_1 \| P'_2$. By applying the induction hypothesis on the premises $P_1 \xrightarrow{\bar{a}\tilde{z}} P'_1$ and $P_2 \xrightarrow{a(\tilde{z})} P'_2$, we have that $( P_1 )_c \to^* \simeq ( P'_1 )_c$ and $( P_2 )_c \to^* \simeq ( P'_2 )_c$. We can apply rule CoML.

$$\frac{[(P_1)_p] \xrightarrow{\{\} \triangleright \overline{\mathsf{tt}}(a,\tilde{z})} [(P'_1)_p] \quad [(P_2)_p] \xrightarrow{\{\} \triangleright \overline{\mathsf{tt}}(a,\tilde{z})} [(P'_2)_p]}{[(P_1)_p] \| [(P_2)_p] \xrightarrow{\{\} \triangleright \overline{\mathsf{tt}}(a,\tilde{z})} [(P'_1)_p] \| [(P'_2)_p]}$$

Now, it is easy to see that: $(P'_1 \| P'_2)_c \simeq [(P'_1)_p] \| [(P'_2)_p]$. Notice that the $b\pi$ term and its encoding have the same observable behavior i.e., $P_1 \| P_2 \downarrow_a$ and $(P_1 \| P_2)_c \downarrow_{(a=a)}$. □

**Lemma 6.2** (Operational Soundness). *If $(P)_c \to Q$ then $\exists P'$ such that $P \to^*_{b\pi} P'$ and $Q \to^* \simeq (P')_c$.*

*Proof.* The proof holds immediately due to the fact that every encoded $b\pi$-term (i.e., $( P )_c$) has exactly one possible transition which matches the original $b\pi$-term (i.e., $P$). □

The idea that we can mimic each transition of $b\pi$-calculus by exactly one transition in $AbC$ implies that soundness and completeness of the operational correspondence can be even proved in a stronger way as in corollary 1 and 2.

**Corollary 6.1** (Strong Completeness). *if $P \to_{b\pi} P'$ then $\exists Q$ such that $Q \equiv (P')_c$ and $(P)_c \to Q$.*

**Corollary 6.2** (Strong Soundness). *if $(\![P]\!)_c \to Q$ then $Q \equiv (\![P']\!)_c$ and $P \to_{b\pi} P'$*

**Theorem 6.1.** *The encoding $(\![\,.\,]\!) : b\pi \to AbC$ is uniform.*

*Proof.* Definition 6.2(1) and 6.2(2) hold by construction. Definition 6.2(4) holds by Lemma 6.1, Lemma 6.2, Corollary 6.1, and Corollary 6.2 respectively. Definition 6.2(3) holds easily and as a result of the proof of Lemma 6.1 and the strong formulation of operational correspondence in Corollary 6.1, and Corollary 6.2, this encoding preserves the observable behavior and cannot introduce divergence. □

As a result of Theorem 4.2, Theorem 6.1 and of the strong formulations of Corollary 6.1, and Corollary 6.2, this encoding is sound and complete with respect to bisimilarity as stated in the following corollaries.

**Corollary 6.3** (Soundness w.r.t bisimilarity).

- $(\![P]\!)_c \approx (\![Q]\!)_c$ *implies* $P \approx Q$

**Corollary 6.4** (Completeness w.r.t bisimilarity).

- $P \approx Q$ *implies* $(\![P]\!)_c \approx (\![Q]\!)_c$

## 7. Concluding Remarks, Related and Future Works

We have introduced a foundational process calculus, named *AbC*, for modeling interactions in CAS systems by relying on attribute-based communication. We tested the expressive power of *AbC* by discussing how other interaction paradigms, such as group based communication and publish-subscribe, could be modelled in our calculus. Moreover, we used *AbC* as the target of the encoding of a process algebra (inspired by CBS [6] and [7]) and proved the correctness of the encoding up to our equivalence.

Now we would like to briefly discuss related works concerning languages and calculi with primitives that either model multiparty interaction or provide interesting ways to establish interaction.

*AbC* is inspired by the SCEL language [18, 19] that was designed to support programming of autonomic computing systems [20]. Compared with SCEL, the knowledge representation in *AbC* is abstract and is not designed for detailed reasoning during systems evolution. This reflects the different objectives of SCEL and *AbC*; SCEL focuses on programming issues, while *AbC* concentrates on introducing a minimal set of primitives to study attribute-based communication.

Many calculi that aim at providing tools for specifying and reasoning about communicating systems have been proposed: CBS [21] captures the essential features of broadcast communication in a simple and natural way. Whenever a process transmits a value, all processes running in parallel, and ready to perform an input, do catch the broadcast. In [22], an LTS for CBS was proposed where notions of strong and weak labeled bisimilarity relying on a discard relation were defined.

The $b\pi$−calculus [7] equips $\pi$−calculus [23] with broadcast primitives where only agents listening on a specific channel can receive the broadcast. The authors also proposed an LTS relying on a discard relation and a labeled bisimilarity which is proved to coincide with the reduction barbed congruence when closed under substitutions. The CPC calculus [24] relies on pattern-matching. Input and output prefixes are generalized to patterns whose unification enables a two-way, or symmetric, flow of information and partners are selected by matching inputs with outputs and testing for equality. The attribute $\pi$-calculus [25] aims at constraining interaction by considering values of communication attributes. A $\lambda$-function is associated to each receiving action and communication takes place only if the result of the evaluation of the function with the provided input falls within a predefined set of values. The imperative $\pi$-calculus [26] is a recent extension of the attribute $\pi$-calculus with a global store and with imperative programs used to specify constraints. The broadcast Quality Calculus of [27] deals with the problem of denial-of-service by means of *selective* input actions. It inspects the structure of messages by associating specific contracts to inputs, but does not provide any mean to change the input contracts during execution.

*AbC* combines the lessons learnt from the above mentioned languages and calculi in the sense that *AbC* strives for expressivity while preserving minimality and simplicity. The dynamic settings of attributes and the possibility of inspecting/modifying the environment gives $AbC$ greater flexibility and expressivity while keeping models as much natural as possible.

We plan to investigate the impact of alternative behavioral relations like testing preorders in terms of equational laws, proof techniques, etc. We want to devise an appropriate notion of temporal logic that can be used to specify, verify, and monitor collective adaptive case studies, modeled in $AbC$. Actually since CAS components usually operate in an open and changing environment, the spatial and temporal dimensions are strictly related and influence each other. Thus, we would like to investigate the impact of spatio-temporal logic approaches in the context of $AbC$ models. One promising approach is presented in [28].

Another line of research worth investigating is anonymity at the level of attribute identifiers. Clearly, $AbC$ achieves dynamicity and openness in the distributed settings, which is an advantage compared to channel-based models. In our model, components are anonymous; however the "name-dependency" challenge arises at another level, that is, the level of attribute environments. In other words, the sender's predicate should be aware of the identifiers of receiver's attributes in order to explicitly use them. For instance, the sending predicate ($loc = (1, 4)$) targets the components at location $(1, 4)$. However, different components might use different identifiers names (i.e., "location") to denote their locations; this requires that there should be an agreement about the attribute identifiers used by the components. For this reason, appropriate mechanisms for handling *attribute directories* together with identifiers matching/correspondence will be considered. These mechanisms will be particularly useful when integrating heterogeneous applications.

Another research direction is introducing a static semantics for $AbC$ as a way to discipline the interaction between components. This way we can answer questions regarding deadlock freedom and if the message is of the expected type for the receiver.

# Appendix A. Proofs

*Appendix A.1. Proof of Lemma 3.1*

We prove each statement separately.

1. **We need to prove that for any $\lambda$ such that $\lambda = \Gamma' \triangleright \Pi(\tilde{v})$ and $\Pi \simeq \text{ff}$, then $C \xrightarrow{\lambda} C$.** We proceed by induction on the syntax of $C$.

   *Base of Induction:* $C = \Gamma :_I P$. It is sufficient to prove that $\Gamma :_I P \xmapsto{\widetilde{\Gamma' \triangleright \Pi(\tilde{v})}} \Gamma :_I P$ where $\Pi \simeq \text{ff}$. This can be done by induction on the transition $\Gamma :_I P \xmapsto{\widetilde{\Gamma' \triangleright \Pi(\tilde{v})}} \hat{C}$ where $\Pi \simeq \text{ff}$. We have the following cases.

   **Case 1:** $P = 0$. We can only apply rule FZERO regardless of $\Pi$ and we have that $\Gamma :_I 0 \xmapsto{\widetilde{\Gamma' \triangleright \Pi(\tilde{v})}} \Gamma :_I 0$ as required.

   **Case 2:** $P = \Pi_1(\tilde{x}).U$. We can only apply rule FRCV because $\Gamma \downarrow I \not\models \Pi$ (Notice that $\Pi \simeq \text{ff}$) and we have that $\Gamma :_I \Pi_1(\tilde{x}).U \xmapsto{\widetilde{\Gamma' \triangleright \Pi(\tilde{v})}} \Gamma :_I \Pi_1(\tilde{x}).U$ as required.

   **Case 3:** $P = (\tilde{E})@\Pi_1.U$. We can only apply rule FBRD regardless of $\Pi$ and we have that $\Gamma :_I (\tilde{E})@\Pi_1.U \xmapsto{\widetilde{\Gamma' \triangleright \Pi(\tilde{v})}} \Gamma :_I (\tilde{E})@\Pi_1.U$ as required.

   **Case 4:** $P = \langle \Pi_1 \rangle P$. We can either apply rule FAWARE1 if $\Gamma \not\models \Pi_1$ or rule FAWARE2 otherwise by the induction hypothesis on the premise $\Gamma :_I P \xmapsto{\widetilde{\Gamma' \triangleright \Pi(\tilde{v})}} \Gamma :_I P$ of rule FAWARE2 and in both cases we have that $\Gamma :_I \langle \Pi_1 \rangle P \xmapsto{\widetilde{\Gamma' \triangleright \Pi(\tilde{v})}} \Gamma :_I \langle \Pi_1 \rangle P$ as required.

   **Case 5:** $P = P_1 + P_2$. We can only apply rule FSUM by the induction hypothesis on its premises $\Gamma :_I P_1 \xmapsto{\widetilde{\Gamma' \triangleright \Pi(\tilde{v})}} \Gamma :_I P_1$ and $\Gamma :_I P_2 \xmapsto{\widetilde{\Gamma' \triangleright \Pi(\tilde{v})}} \Gamma :_I P_2$ and we have that $\Gamma :_I P_1 + P_2 \xmapsto{\widetilde{\Gamma' \triangleright \Pi(\tilde{v})}} \Gamma :_I P_1 + P_2$ as required.

   **Case 6:** $P_1 | P_2$. It is proved in a similar case of Case 5 by applying rule FINT instead.

   **Case 7:** $P = K$. It is proved by applying rule FREC and by the induction hypothesis on the premise $\Gamma :_I P \xmapsto{\widetilde{\Gamma' \triangleright \Pi(\tilde{v})}} \Gamma :_I P$ of FREC.

   Hence, we have that $\Gamma :_I P \xmapsto{\widetilde{\Gamma' \triangleright \Pi(\tilde{v})}} \Gamma :_I P$ where $\Pi \simeq \text{ff}$. By applying rule FCOMP, we have that $\Gamma :_I P \xrightarrow{\Gamma' \triangleright \Pi(\tilde{v})} \Gamma :_I P$ when $\Pi \simeq \text{ff}$.

   *Inductive Hypothesis:.* Let us assume that for any $C_1$ and $C_2$, and for any $\lambda$ such that $\lambda = \Gamma' \triangleright \Pi(\tilde{v})$ and $\Pi \simeq \text{ff}$, then $C_i \xrightarrow{\lambda} C_i$

   *Inductive Step:.* We have to consider three cases: $C = C_1 \| C_2$, $C = [\ C_i\ ]^{\triangleright f}$ and $C = [\ C_i\ ]^{\triangleleft f}$.

   **Case 1:** $C = C_1 \| C_2$. The statement follows directly from the inductive hypothesis by applying rule SYNC:
   $$\frac{C_1 \xrightarrow{\lambda} C_1 \quad C_2 \xrightarrow{\lambda} C_2}{C_1 \| C_2 \xrightarrow{\lambda} C_1 \| C_2}$$

   **Case 2:** $C = [\ C_i\ ]^{\triangleright f}$. This case follows from the inductive hypothesis ($C_i \xrightarrow{\lambda} C_i$) by applying rule RESO and by observing that if $\Pi \simeq \text{ff}$ then for any $\Pi'$, $\Pi \wedge \Pi' \simeq \text{ff}$.

   **Case 3:** $C = [\ C_i\ ]^{\triangleleft f}$. Exactly in the previous case by considering rule RESI.

2. **We need to prove that if $C_1 \xrightarrow{\lambda} C_1'$ and $\lambda = \tau$, then $C_1 \| C \xrightarrow{\tau} C_1' \| C$ and $C \| C_1 \xrightarrow{\tau} C \| C_1'$.** The statement follows directly from the previous point by applying rules CoML and CoMR.

3. **We need to prove that if $C_1 \Rightarrow C_1'$ then $C_1 \| C \Rightarrow C_1' \| C$ and $C \| C_1 \Rightarrow C \| C_1'$.** We prove the statement ($C_1 \Rightarrow C_1'$ then $C_1 \| C \Rightarrow C_1' \| C$) and the statement ($C_1 \Rightarrow C_1'$ then $C \| C_1 \Rightarrow C \| C_1'$) can be proved in a symmetric way. The proof proceeds by induction on the length of the derivation $\Rightarrow_w$.

    **Base case, $w = 0$:** We have that $C_1 \Rightarrow_0 C_1$ and $C_1 \| C \Rightarrow_0 C_1 \| C$ as required.

    **Inductive Hypothesis:** we assume that $\forall k \leq w : C_1 \Rightarrow_k C_1'$ then $C_1 \| C \Rightarrow_k C_1' \| C$.

    **Inductive Step:** Let $C_1 \Rightarrow_{w+1} C_1'$. By definition of $\Rightarrow_{w+1}$, we have that there exists $C_1''$ such that $C_1 \xrightarrow{\tau} C_1''$ and $C_1'' \Rightarrow_w C_1'$. By the second statement of this lemma, we have that if $C_1 \xrightarrow{\tau} C_1''$ then $C_1 \| C \xrightarrow{\tau} C_1'' \| C$. Moreover, by the induction hypothesis $C_1'' \| C \Rightarrow_w C_1' \| C$. Hence, we have that $C_1 \Rightarrow C_1'$ then $C_1 \| C \Rightarrow C_1' \| C$ as required.

4. **We need to prove that if $C_1 \xRightarrow{\Gamma \triangleright \Pi_1(\tilde{v})} C_1'$ and $\Pi_1 \simeq \Pi_2$ then $C_1 \xRightarrow{\Gamma \triangleright \Pi_2(\tilde{v})} C_1'$** We first need to prove the single-step version of this statement: if $C_1 \xrightarrow{\Gamma \triangleright \Pi_1(\tilde{v})} C_1'$ and $\Pi_1 \simeq \Pi_2$ then $C_1 \xrightarrow{\Gamma \triangleright \Pi_2(\tilde{v})} C_1'$. The proof proceeds by induction on $C_1$.

    **Case 1:** $C_1 = \Gamma :_I P$. We have to proceed by induction on the length of the derivation of the transitions $\Gamma :_I P \xrightarrow{\Gamma' \triangleright \Pi_1(\tilde{v})} \Gamma' :_I P'$ and $\Gamma :_I P \xmapsto{\widetilde{\Gamma' \triangleright \Pi_1(\tilde{v})}} \Gamma :_I P$.

    - We start by the transition $\Gamma :_I P \xmapsto{\widetilde{\Gamma' \triangleright \Pi_1(\tilde{v})}} \Gamma :_I P$. The cases when rules FBRD, FAWARE1, FZERO are trivial since they refuse regardless of the sender predicate. The other discard rules can be proved by the induction hypothesis on their premises. The interesting case is when rule FRCV is applied. In this case $P = \Pi(\tilde{x}).U$ and we have that $\Gamma :_I \Pi(\tilde{x}).U \xmapsto{\widetilde{\Gamma' \triangleright \Pi_1(\tilde{v})}} \Gamma :_I \Pi(\tilde{x}).U$ only if $\Gamma' \not\models \{\Pi[\tilde{v}/\tilde{x}]\}_\Gamma$ or $\Gamma \downarrow I \not\models \Pi_1$. Now we need to show that $\Gamma \downarrow I \not\models \Pi_2$, but this is immediate from Definition 2.1 and the fact that $\Pi_1 \simeq \Pi_2$ and we have that $\Gamma :_I \Pi(\tilde{x}).U \xmapsto{\widetilde{\Gamma' \triangleright \Pi_2(\tilde{v})}} \Gamma :_I \Pi(\tilde{x}).U$. By applying rule FCOMP, we have that we have that $\Gamma :_I P \xrightarrow{\Gamma' \triangleright \Pi_1(\tilde{v})} \Gamma :_I P$ implies $\Gamma :_I P \xrightarrow{\Gamma' \triangleright \Pi_2(\tilde{v})} \Gamma :_I P$ such that $\Pi_1 \simeq \Pi_2$ as required.

    - Now we prove the transition $\Gamma :_I P \xrightarrow{\Gamma' \triangleright \Pi_1(\tilde{v})} \Gamma' :_I P'$. The interesting case is when rule RCV is applied and other cases are proved by the induction hypothesis on their premises. In this case $P = \Pi(\tilde{x}).U$ and we have that $\Gamma :_I \Pi(\tilde{x}).U \xrightarrow{\Gamma' \triangleright \Pi_1(\tilde{v})} \{\!|\Gamma :_I U[\tilde{v}/\tilde{x}]|\!\}$ only if $\Gamma' \models \{\Pi[\tilde{v}/\tilde{x}]\}_\Gamma$ or $\Gamma \downarrow I \models \Pi_1$. Now we need to show that $\Gamma \downarrow I \models \Pi_2$, but this is immediate from Definition 2.1 and the fact that $\Pi_1 \simeq \Pi_2$ and we have that $\Gamma :_I \Pi(\tilde{x}).U \xrightarrow{\Gamma' \triangleright \Pi_2(\tilde{v})} \{\!|\Gamma :_I U[\tilde{v}/\tilde{x}]|\!\}$. By applying rule COMP, we have that we have that $\Gamma :_I P \xrightarrow{\Gamma' \triangleright \Pi_1(\tilde{v})} \Gamma' :_I P'$ implies $\Gamma :_I P \xrightarrow{\Gamma' \triangleright \Pi_2(\tilde{v})} \Gamma' :_I P'$ such that $\Pi_1 \simeq \Pi_2$ as required.

    Hence, we have that if $\Gamma :_I P \xrightarrow{\Gamma \triangleright \Pi_1(\tilde{v})} \Gamma' :_I P'$ and $\Pi_1 \simeq \Pi_2$ then $\Gamma :_I P \xrightarrow{\Gamma \triangleright \Pi_2(\tilde{v})} \Gamma' :_I P'$ as required.

    **Case 2:** $C = C_3 \| C_4$. We can only use rule SYNC and by the induction hypothesis on the premises of SYNC to prove this case.

    **Case 3:** $C = [\ C_3\ ]^{\triangleleft f}$ or $C = [\ C_3\ ]^{\triangleright f}$. The statement follows directly from the inductive hypothesis by applying rules RESO and RESI and from the fact that $\Pi_1 \simeq \Pi_2$ implies that $\Pi_1 \wedge \Pi \simeq \Pi_2 \wedge \Pi$ for any predicate $\Pi$.

Let $C_1 \xRightarrow{\Gamma \triangleright \Pi_1(\tilde{v})} C_1'$. This means that there exits $C_2$ and $C_2'$ such that:

$$C_1 \Rightarrow C_2 \xrightarrow{\Gamma \triangleright \Pi_1(\tilde{v})} C_2' \Rightarrow C_1'$$

We have already proved that $C_2 \xrightarrow{\Gamma \triangleright \Pi_2(\tilde{v})} C_2'$ for any $\Pi_2 \simeq \Pi_1$. Hence:

$$C_1 \Rightarrow C_2 \xrightarrow{\Gamma \triangleright \Pi_2(\tilde{v})} C_2' \Rightarrow C_1'$$

from which we can finally infer that $C_1 \xRightarrow{\Gamma \triangleright \Pi_2(\tilde{v})} C_1'$.

5. **We prove that if $C_1 \xrightarrow{\tau} C_1'$, then for any $f$:** $[\,C_1\,]^{\triangleright f} \xrightarrow{\tau} [\,C_1'\,]^{\triangleright f}$ **and** $[\,C_1\,]^{\triangleleft f} \xrightarrow{\tau} [\,C_1'\,]^{\triangleleft f}$. If $C_1 \xrightarrow{\tau} C_1'$ then there exists $\Gamma$, $\tilde{v}$ and $\Pi \simeq \mathsf{ff}$ such that $C_1 \xrightarrow{\Gamma \triangleright \overline{\Pi}(\tilde{v})} C_1'$.

    By applying rule ResO we have that for any $f$, $[\,C_1\,]^{\triangleright f} \xrightarrow{\Gamma \triangleright \overline{\Pi \wedge \Pi'}(\tilde{v})} [\,C_1'\,]^{\triangleright f}$ where $f(\Gamma, \tilde{v}) = \Pi'$. Since, $\Pi \simeq \mathsf{ff}$ we have that $\Pi \wedge \Pi' \simeq \mathsf{ff} \wedge \Pi' \simeq \mathsf{ff}$. Hence, $[\,C_1\,]^{\triangleright f} \xrightarrow{\tau} [\,C_1'\,]^{\triangleright f}$

    We can also apply rule ResIPass to prove that $[\,C_1\,]^{\triangleleft f} \xrightarrow{\Gamma \triangleright \overline{\Pi}(\tilde{v})} [\,C_1'\,]^{\triangleleft f}$. That is $[\,C_1\,]^{\triangleleft f} \xrightarrow{\tau} [\,C_1'\,]^{\triangleleft f}$

6. **We prove that if $C_1 \Rightarrow C_1'$, then for any $f$:** $[\,C_1\,]^{\triangleright f} \Rightarrow [\,C_1'\,]^{\triangleright f}$ **and** $[\,C_1\,]^{\triangleleft f} \Rightarrow [\,C_1'\,]^{\triangleleft f}$. We prove the statement by induction on $C_1 \Rightarrow_w C_1'$.

    **Base case, $w = 0$:** We have that $C_1 \Rightarrow_0 C_1$ while both $[\,C_1\,]^{\triangleright f} \Rightarrow_0 [\,C_1\,]^{\triangleright f}$ and $[\,C_1\,]^{\triangleleft f} \Rightarrow_0 [\,C_1\,]^{\triangleleft f}$ as required.

    **Inductive Hypothesis:** we assume that $\forall k \leq w : C_1 \Rightarrow_k C_1'$ then: $[\,C_1\,]^{\triangleright f} \Rightarrow_k [\,C_1'\,]^{\triangleright f}$ and $[\,C_1\,]^{\triangleleft f} \Rightarrow_k [\,C_1'\,]^{\triangleleft f}$.

    **Inductive Step:** Let $C_1 \Rightarrow_{w+1} C_1'$. By definition of $\Rightarrow_{w+1}$, we have that there exists $C_1''$ such that $C_1 \xrightarrow{\tau} C_1''$ and $C_1'' \Rightarrow_w C_1'$. By **item 5** of this Lemma, and by Inductive Hypothesis we have that for any $f$:

    - $[\,C_1\,]^{\triangleright f} \xrightarrow{\tau} [\,C_1''\,]^{\triangleright f}$ and $[\,C_1''\,]^{\triangleright f} \Rightarrow_w [\,C_1'\,]^{\triangleright f}$;
    - $[\,C_1\,]^{\triangleleft f} \xrightarrow{\tau} [\,C_1''\,]^{\triangleleft f}$ and $[\,C_1''\,]^{\triangleleft f} \Rightarrow_w [\,C_1'\,]^{\triangleleft f}$.

    From the two above we have that $[\,C_1\,]^{\triangleright f} \Rightarrow_{w+1} [\,C_1'\,]^{\triangleright f}$ and $[\,C_1\,]^{\triangleleft f} \Rightarrow_{w+1} [\,C_1'\,]^{\triangleleft f}$.

*Appendix A.2. Proof of Lemma 4.2*

It is sufficient to prove that the relation

$$\mathcal{R} = \{(C_1 \| C, C_2 \| C) | \forall C_1, C_2, C \in \mathrm{Comp} : C_1 \approx C_2\}$$

is a weak bisimulation. First of all we can observe that $\mathcal{R}$ is *symmetric*. It is easy to see that if $(C_1 \| C, C_2 \| C) \in \mathcal{R}$ then $(C_2 \| C, C_1 \| C) \in \mathcal{R}$. We have now to prove that for each $(C_1 \| C, C_2 \| C) \in \mathcal{R}$ and for each $\lambda_1$ such that $bn(\lambda_1) \cap fn(C_1, C_2) = \emptyset$:

$$C_1 \| C \xrightarrow{\lambda_1} C_3 \text{ implies } \exists \lambda_2 : \lambda_1 \simeq \lambda_2 \text{ such that } C_2 \| C \xRightarrow{\widehat{\lambda_2}} C_4 \text{ and } (C_3, C_4) \in \mathcal{R}.$$

We can observe that the transition $C_1 \| C \xrightarrow{\lambda_1} C_3$ can be derived by using one of rule among SYNC, COML, and COMR. The following cases can be distinguished:

**Case 1: rule Sync is applied.** In this case $\lambda_1 = \Gamma \triangleright \Pi_1(\tilde{v})$, $C_1 \xrightarrow{\Gamma \triangleright \Pi_1(\tilde{v})} C_1'$, $C \xrightarrow{\Gamma \triangleright \Pi_1(\tilde{v})} C'$ and $C_3 = C_1' \| C'$. Since $C_1 \approx C_2$, we have that $C_2 \xRightarrow{\lambda_2} C_2'$, with $\lambda_1 \simeq \lambda_2 = \Gamma \triangleright \Pi_2(\tilde{v})$ and $C_1' \approx C_2'$. This implies that there exists $C_2''$ and $C_2'''$ such that $C_2 \Rightarrow C_2'' \xrightarrow{\lambda_2} C_2''' \Rightarrow C_2'$. Moreover, since $\lambda_1 \simeq \lambda_2$, by Lemma 3.1, we have that $C \xrightarrow{\lambda_2} C'$. By using again Lemma 3.1, we have that $C_2 \| C \Rightarrow C_2'' \| C$. We can apply rule SYNC to prove that $C_2'' \| C \xrightarrow{\lambda_2} C_2''' \| C'$. Finally, as before, we have that $C_2''' \| C' \Rightarrow C_2' \| C'$. The statement follows by observing that $(C_1' \| C', C_2' \| C') \in \mathcal{R}$.

**Case 2: rule ComL is applied.** We can distinguish two cases: $\lambda_1 = \tau$, $\lambda_1 \neq \tau$. If $\lambda_1$ is a *silent* transition ($\lambda_1 = \tau$), we have that $\lambda_1 = \Gamma \triangleright \overline{\Pi_1}(\tilde{v})$ with $\Pi_1 \simeq \text{ff}$. Moreover, $C_1 \xrightarrow{\lambda_1} C_1'$ and $C_3 = C_1' \| C$ (since $\Pi_1 \simeq \text{ff}$, $C \xrightarrow{\Gamma \triangleright \Pi_1(\tilde{v})} C$). From the fact that $C_1 \approx C_2$, we have that $C_2 \Rightarrow C_2'$ and $C_1' \approx C_2'$. Moreover, by Lemma 3.1, we have that $C_2 \| C \Rightarrow C_2' \| C$. It is easy to observe that $(C_1' \| C, C_2' \| C) \in \mathcal{R}$.

If $\lambda_1 \neq \tau$, then $\lambda_1 = \Gamma \triangleright \overline{\Pi_1}(\tilde{v})$. We have that $C_1 \xrightarrow{\Gamma \triangleright \overline{\Pi_1}(\tilde{v})} C_1'$, $C \xrightarrow{\Gamma \triangleright \Pi_1(\tilde{v})} C'$ and $C_3 = C_1' \| C'$. In this case the statement follows similarly to **Case 1**. Indeed, $C_2 \xRightarrow{\Gamma \triangleright \overline{\Pi_2}(\tilde{v})} C_2'$, with $\Gamma \triangleright \overline{\Pi_1}(\tilde{v}) \simeq \Gamma \triangleright \overline{\Pi_2}(\tilde{v})$ and $C_1' \approx C_2'$. Moreover, by Lemma 3.1, $C \xrightarrow{\Gamma \triangleright \Pi_2(\tilde{v})} C'$. Hence, $C_2 \| C \xRightarrow{\Gamma \triangleright \overline{\Pi_2}(\tilde{v})} C_2' \| C'$ and $(C_1' \| C', C_2' \| C') \in \mathcal{R}$.

**Case 3: rule ComR is applied.** We can distinguish two cases: $\lambda_1 = \tau$, $\lambda_1 \neq \tau$. By Lemma 3.1 we have that $C \xrightarrow{\lambda_1} C'$ and $C_3 = C_1 \| C'$. Similarly, $C_2 \| C \xRightarrow{\lambda_1} C_2 \| C'$ and $(C_1 \| C', C_2 \| C') \in \mathcal{R}$. If $\lambda_1 \neq \tau$ we have that $\lambda_1 = \Gamma \triangleright \overline{\Pi_1}(\tilde{v})$, $C_1 \xrightarrow{\Gamma \triangleright \Pi_1(\tilde{v})} C_1'$, $C \xrightarrow{\Gamma \triangleright \overline{\Pi_1}(\tilde{v})} C'$, and $C_3 = C_1' \| C'$. Like in the previous cases, by using the fact that $C_1 \approx C_2$ and by Lemma 3.1, we have that $C_2 \xRightarrow{\Gamma \triangleright \Pi_1(\tilde{v})} C_2'$ and $C_1' \approx C_2'$, and that $C_2 \| C' \xRightarrow{\lambda_1} C_2' \| C'$. The statement follows by observing that also in this case $(C_1' \| C', C_2' \| C') \in \mathcal{R}$.

The strong case of bisimulation ($\sim$) follows in a similar way.

*Proof of Lemma 4.3.* We prove the lemma case by case. We start by the output restriction operator and we follow up with the input restriction one. For the first case, It is sufficient to prove that the relation

$$\mathcal{R} = \{([\ C_1\ ]^{\triangleright f}, [\ C_2\ ]^{\triangleright f}) | \text{ for all functions } f \text{ and all } C_1, C_2 \in \text{Comp} : C_1 \approx C_2\}$$

is a weak bisimulation. First of all we can observe that $\mathcal{R}$ is *symmetric*. We have now to prove that for each $([\ C_1\ ]^{\triangleright f}, [\ C_2\ ]^{\triangleright f}) \in \mathcal{R}$ and for each $\lambda_1$:

$$[\ C_1\ ]^{\triangleright f} \xrightarrow{\lambda_1} C_3 \text{ implies } \exists \lambda_2 : \lambda_1 \simeq \lambda_2 \text{ such that } [\ C_2\ ]^{\triangleright f} \xRightarrow{\widehat{\lambda_2}} C_4 \text{ and } (C_3, C_4) \in \mathcal{R}.$$

We can observe that the transition $[\ C_1\ ]^{\triangleright f} \xrightarrow{\lambda_1} C_3$ can be derived either by using rule RESO or by using RESOPASS.

When rule RESO is used, $\lambda_1 = \Gamma \triangleright \overline{\Pi_1 \wedge \Pi}(\tilde{v})$ where $f(\Gamma, \tilde{v}) = \Pi$ and $C_1 \xrightarrow{\Gamma \triangleright \overline{\Pi_1}(\tilde{v})} C_1'$. Since $C_1 \approx C_2$, we have that $C_2 \xRightarrow{\Gamma \triangleright \overline{\Pi_2}(\tilde{v})} C_2'$, with $\Gamma \triangleright \overline{\Pi_1}(\tilde{v}) \simeq \Gamma \triangleright \overline{\Pi_2}(\tilde{v})$ and $C_1' \approx C_2'$. From the latter we have that, by definition of $\mathcal{R}$, $([\ C_1'\ ]^{\triangleright f}, [\ C_2'\ ]^{\triangleright f}) \in \mathcal{R}$.

We can now proceed by considering two cases: $\Pi_1 \simeq \text{ff}$, $\Pi_1 \neq \text{ff}$.

If $\Pi_1 \simeq \text{ff}$, we have that $C_2 \Rightarrow C_2'$. Directly from *Item 6* of Lemma 3.1 we have that $[\ C_2\ ]^{\triangleright f} \Rightarrow [\ C_2'\ ]^{\triangleright f}$.

If $\Pi_1 \neq \text{ff}$, we have that there exists $C_2''$ and $C_2'''$ such that $C_2 \Rightarrow C_2'' \xrightarrow{\Gamma \triangleright \overline{\Pi_2}(\tilde{v})} C_2''' \Rightarrow C_2'$. We have that:

- From *Item 6* of Lemma 3.1, $[\ C_2\ ]^{\triangleright f} \Rightarrow [\ C_2''\ ]^{\triangleright f}$;

- By applying rule RESO, $[\ C_2''\ ]^{\triangleright f} \xrightarrow{\lambda_2} [\ C_2'''\ ]^{\triangleright f}$ where $\lambda_1 \simeq \lambda_2 = \Gamma \triangleright \overline{\Pi_2 \wedge \Pi}(\tilde{v})$;

- From *Item 6* of Lemma 3.1, $[\ C_2'''\ ]^{\triangleright f} \Rightarrow [\ C_2'\ ]^{\triangleright f}$.

That is, $[\ C_2\ ]^{\triangleright f} \xRightarrow{\lambda_2} [\ C_2'\ ]^{\triangleright f}$.

When rule RESOPASS is used, we have that $\lambda_1 = \Gamma \triangleright \Pi_1(\tilde{v})$, $C_1 \xrightarrow{\lambda_1} C_1'$ and $C_3 = [\ C_1'\ ]^{\triangleright f}$. Since $C_1 \approx C_2$, we have that $C_2 \xRightarrow{\Gamma \triangleright \Pi_2(\tilde{v})} C_2'$ and $C_1' \approx C_2'$ (and $([\ C_1'\ ]^{\triangleleft f}, [\ C_2'\ ]^{\triangleleft f}) \in \mathcal{R}$).

Hence we have that, $C_2 \Rightarrow C_2'' \xrightarrow{\Gamma \triangleright \Pi_2(\tilde{v})} C_2''' \Rightarrow C_2'$

By using Lemma 3.1, and by applying rule RULEOPASS, we have that $[\ C_2\ ]^{\triangleright f} \xRightarrow{\Gamma \triangleright \Pi_2(\tilde{v})} [\ C_2'\ ]^{\triangleright f}$, and the statement follows.

For the second case, It is sufficient to prove that the relation

$$\mathcal{R} = \{([\ C_1\ ]^{\triangleleft f}, [\ C_2\ ]^{\triangleleft f}) |\ \text{for all functions}\ f\ \text{and all}\ C_1, C_2 \in \text{Comp} : C_1 \approx C_2\}$$

is a weak bisimulation. First of all we can observe that $\mathcal{R}$ is *symmetric*. We have now to prove that for each $([\ C_1\ ]^{\triangleleft f}, [\ C_2\ ]^{\triangleleft f}) \in \mathcal{R}$ and for each $\lambda_1$:

$$[\ C_1\ ]^{\triangleleft f} \xrightarrow{\lambda_1} C_3\ \text{implies}\ \exists \lambda_2 : \lambda_1 \simeq \lambda_2\ \text{such that}\ [\ C_2\ ]^{\triangleleft f} \xRightarrow{\widehat{\lambda_2}} C_4\ \text{and}\ (C_3, C_4) \in \mathcal{R}.$$

We can observe that the transition $[\ C_1\ ]^{\triangleleft f} \xrightarrow{\lambda_1} C_3$ can be derived by using either rule ResI or ResIPass.

If rule ResI is used, we have that $\lambda_1 = \Gamma \triangleright \Pi_1(\tilde{v})$ where $f(\Gamma, \tilde{v}) = \Pi$ and $C_1 \xrightarrow{\Gamma \triangleright (\Pi_1 \wedge \Pi)(\tilde{v})} C_1'$. Since $C_1 \approx C_2$, we have that $C_2 \xRightarrow{\Gamma \triangleright (\Pi_2 \wedge \Pi)(\tilde{v})} C_2'$, with $\Pi_1 \simeq \Pi_2$ and $C_1' \approx C_2'$. This implies that there exists $C_2''$ and $C_3'''$ such that $C_2 \Rightarrow C_2'' \xrightarrow{\Gamma \triangleright (\Pi_2 \wedge \Pi)(\tilde{v})} C_2''' \Rightarrow C_2'$. By using again Lemma 3.1, we have that $[\ C_2\ ]^{\triangleleft f} \Rightarrow [\ C_2''\ ]^{\triangleleft f}$. We can apply rule ResI to prove that $[\ C_2''\ ]^{\triangleleft f} \xrightarrow{\lambda_2} [\ C_2'''\ ]^{\triangleleft f}$ with $\lambda_2 = \Gamma \triangleright \Pi_2(\tilde{v})$. Finally, as before, we have that $[\ C_2'''\ ]^{\triangleleft f} \Rightarrow [\ C_2'\ ]^{\triangleleft f}$. The statement follows by observing that $([\ C_1'\ ]^{\triangleleft f}, [\ C_2'\ ]^{\triangleleft f}) \in \mathcal{R}$.

When rule ResIPass is used, we have that $\lambda_1 = \Gamma \triangleright \overline{\Pi_1}(\tilde{v})$, $C_1 \xrightarrow{\lambda_1} C_1'$ and $C_3 = [\ C_1'\ ]^{\triangleleft f}$. Since $C_1 \approx C_2$, we have that $C_2 \xRightarrow{\Gamma \triangleright \overline{\Pi_2}(\tilde{v})} C_2'$ and $C_1' \approx C_2'$. By using Lemma 3.1, and by applying rule RuleIPass, we have that $[\ C_2\ ]^{\triangleleft f} \xRightarrow{\Gamma \triangleright \overline{\Pi_2}(\tilde{v})} [\ C_2'\ ]^{\triangleleft f}$. This case follows directly from the fact that $C_1' \approx C_2'$ and $([\ C_1'\ ]^{\triangleleft f}, [\ C_2'\ ]^{\triangleleft f}) \in \mathcal{R})$.

The strong case of bisimulation ($\sim$) follows in a similar way. $\square$

## Appendix B. Detailed proofs about the encoding

*of Lemma 6.1.* The proof proceeds by induction on the shortest transition of $\rightarrow_{b\pi}$. We have several cases depending on the structure of the term $P$.

- if $P \triangleq \texttt{nil}$: This case is immediate $(\!|\texttt{nil}|\!)_c \triangleq [0]$

- if $P \triangleq \tau.G$: We have that $\tau.G \xrightarrow{\tau} G$ and it is translated to $(\!|\tau.G|\!)_c \triangleq [()@\textsf{ff}.(\!|G|\!)_p]$. We can only apply rule Comp to mimic this transition.

$$\frac{[()@\textsf{ff}.(\!|G|\!)_p] \xmapsto{\{\} \triangleright \overline{\textsf{ff}}()} [(\!|G|\!)_p]}{[()@\textsf{ff}.(\!|G|\!)_p] \xrightarrow{\{\} \triangleright \overline{\textsf{ff}}()} [(\!|G|\!)_p]}$$

  From Table 6, we have that $(\!|\ G\ |\!)_c = [(\!|G|\!)_p]$ as required. Notice that sending on a false predicate is not observable (i.e., a silent move).

- if $P \triangleq a(\tilde{x}).G$: We have that $a(\tilde{x}).G \xrightarrow{a(\tilde{z})} G[\tilde{z}/\tilde{x}]$ and it is translated to $(\!|a(\tilde{x}).Q|\!)_c \triangleq [\Pi(y, \tilde{x}).(\!|G|\!)_p]]$ where $\Pi = (y = a)$. We can only apply rule Comp to mimic this transition.

$$\frac{[\Pi(y, \tilde{x}).(\!|G|\!)_p] \xmapsto{\{\} \triangleright \textsf{tt}(a,\ \tilde{z})} [(\!|G|\!)_p[a/y,\ \tilde{z}/\tilde{x}]]}{[\Pi(y, \tilde{x}).(\!|G|\!)_p] \xrightarrow{\{\} \triangleright \textsf{tt}(a,\ \tilde{z})} [(\!|G|\!)_p[a/y,\ \tilde{z}/\tilde{x}]]}$$

  From Table 6, It is not hard to see that: $(\!|G[\tilde{z}/\tilde{x}]|\!)_c \simeq [(\!|G|\!)_p[a/y,\ \tilde{z}/\tilde{x}]] \simeq [(\!|G|\!)_p[\tilde{z}/\tilde{x}]]$ since $y \notin n((\!|G|\!)_p)$.

- if $P \triangleq \bar{a}\tilde{x}.G$: The proof is similar to the previous case but by applying an output transition instead.

- The fail rules for **nil**, $\tau$, input and output are proved in a similar way but with applying FComp instead.

- if $P \triangleq ((rec\ A\langle\tilde{x}\rangle).P)\langle\tilde{y}\rangle)$: This case is trivial.

- if $P \triangleq G_1 + G_2$: We have that either $G_1 + G_2 \xrightarrow{\alpha} G'_1$ or $G_1 + G_2 \xrightarrow{\alpha} G'_2$. We only consider the first case with $G_1 \xrightarrow{\alpha} G'_1$ and the other case follows in a similar way. This process is translated to $(\!|G_1 + G_2|\!)_c \triangleq [(\!|G_1|\!)_p + (\!|G_2|\!)_p]$. By applying the induction hypothesis on the premise $G_1 \xrightarrow{\alpha} G'_1$, we have that $(\!|G_1|\!)_c \to^* \simeq (\!|G'_1|\!)_c$. We can apply either rule COMP or rule FCOMP (i.e., when discarding) to mimic this transition depending on the performed action. We consider the case of COMP only and the other case follows in a similar way.

$$\frac{[(\!|G_1|\!)_p] \xrightarrow{\lambda} [(\!|G'_1|\!)_p]}{\frac{[(\!|G_1|\!)_p + (\!|G_2|\!)_p] \xrightarrow{\lambda} [(\!|G'_1|\!)_p]}{[(\!|G_1|\!)_p + (\!|G_2|\!)_p] \xrightarrow{\lambda} [(\!|G'_1|\!)_p]}}$$

Again $(\!|G'_1|\!)_c \simeq [\,(\!|G'_1|\!)_p]$

- if $P \triangleq P_1 \| P_2$: This process is translated to $(\!|P_1 \| P_2|\!)_c \triangleq [(\!|P_1|\!)_p] \| [(\!|P_2|\!)_p]$. We have four cases depending on the performed action in deriving the transition $P_1 \| P_2 \xrightarrow{\alpha} \hat{P}$.

  - $P_1 \| P_2 \xrightarrow{\bar{a}\tilde{x}} P'_1 \| P'_2$: We have two cases, either $P_1 \xrightarrow{\bar{a}\tilde{x}} P'_1$ and $P_2 \xrightarrow{a(\tilde{x})} P'_2$ or $P_2 \xrightarrow{\bar{a}\tilde{x}} P'_2$ and $P_1 \xrightarrow{a(\tilde{x})} P'_1$. We only consider the first case and the other case follows in the same way. By applying the induction hypothesis on the premises $P_1 \xrightarrow{\bar{a}\tilde{x}} P'_1$ and $P_2 \xrightarrow{a(\tilde{x})} P'_2$, we have that $(\!|P_1|\!)_c \to^* \simeq (\!|P'_1|\!)_c$ and $(\!|P_2|\!)_c \to^* \simeq (\!|P'_2|\!)_c$. We only can apply COML.

  $$\frac{[(\!|P_1|\!)_p] \xrightarrow{\{\}\triangleright\overline{\text{tt}}(a,\,\tilde{x})} [(\!|P'_1|\!)_p] \quad [(\!|P_2|\!)_p] \xrightarrow{\{\}\triangleright\text{tt}(a,\,\tilde{x})} [(\!|P'_2|\!)_p]}{[(\!|P_1|\!)_p] \| [(\!|P_2|\!)_p] \xrightarrow{\{\}\triangleright\overline{\text{tt}}(a,\,\tilde{x})} [(\!|P'_1|\!)_p] \| [(\!|P'_2|\!)_p]}$$

  Again we have that: $(\!|P'_1 \| P'_2|\!)_c \simeq [(\!|P'_1|\!)_p] \| [(\!|P'_2|\!)_p]$. Notice that the $b\pi$ term and its encoding have the same observable behavior i.e., $P_1 \| P_2 \downarrow_a$ and $(\!|P_1 \| P_2|\!)_c \downarrow_{(\text{tt})}$.

  - $P_1 \| P_2 \xrightarrow{a(\tilde{x})} P'_1 \| P'_2$: By applying the induction hypothesis on the premises $P_1 \xrightarrow{a(\tilde{x})} P'_1$ and $P_2 \xrightarrow{a(\tilde{x})} P'_2$, we have that $(\!|P_1|\!)_c \to^* \simeq (\!|P'_1|\!)_c$ and $(\!|P_2|\!)_c \to^* \simeq (\!|P'_2|\!)_c$. We only can apply SYNC to mimic this transition.

  $$\frac{[(\!|P_1|\!)_p] \xrightarrow{\{\}\triangleright\text{tt}(a,\,\tilde{x})} [(\!|P'_1|\!)_p] \quad [(\!|P_2|\!)_p] \xrightarrow{\{\}\triangleright\text{tt}(a,\,\tilde{x})} [(\!|P'_2|\!)_p]}{[(\!|P_1|\!)_p] \| [(\!|P_2|\!)_p] \xrightarrow{\{\}\triangleright\text{tt}(a,\,\tilde{x})} [(\!|P'_1|\!)_p] \| [(\!|P'_2|\!)_p]}$$

  Again we have that: $(\!|P'_1 \| P'_2|\!)_c \simeq [(\!|P'_1|\!)_p] \| [(\!|P'_2|\!)_p]$.

  - $P_1 \| P_2 \xrightarrow{\alpha} P'_1 \| P_2$ if $P_1 \xrightarrow{\alpha} P'_1$ and $P_2 \xrightarrow{sub(\alpha):}$ or $P_1 \| P_2 \xrightarrow{\alpha} P_1 \| P'_2$ if $P_2 \xrightarrow{\alpha} P'_2$ and $P_1 \xrightarrow{sub(\alpha):}$. we consider only the first case and by applying the induction hypothesis on the premises $P_1 \xrightarrow{\alpha} P'_1$ and $P_2 \xrightarrow{sub(\alpha):}$, we have that $(\!|P_1|\!)_c \to^* \simeq (\!|P'_1|\!)_c$ and $(\!|P_2|\!)_c \to^* \simeq (\!|P_2|\!)_c$. We have many cases depending on the performed action:

  1. if $\alpha = \tau$ then $P_1 \| P_2 \xrightarrow{\tau} P'_1 \| P_2$ with $P_1 \xrightarrow{\tau} P'_1$ and $P_2 \xrightarrow{sub(\tau):}$. We can apply COML to mimic this transition.

  $$\frac{[(\!|P_1|\!)_p] \xrightarrow{\{\}\triangleright\overline{\text{ff}}()} [(\!|P'_1|\!)_p] \quad [(\!|P_2|\!)_p] \xrightarrow{\widetilde{\{\}\triangleright\text{ff}()}} [(\!|P_2|\!)_p]}{[(\!|P_2|\!)_p] \xrightarrow{\{\}\triangleright\text{ff}()} [(\!|P_2|\!)_p]}{[(\!|P_1|\!)_p] \| [(\!|P_2|\!)_p] \xrightarrow{\{\}\triangleright\overline{\text{ff}}()} [(\!|P'_1|\!)_p] \| [(\!|P_2|\!)_p]}$$

and again we have that: $(\!|P_1'\|P_2|\!)_c \simeq [(\!|P_1'|\!)_p] \| \ [(\!|P_2|\!)_p]$.

2. if $\alpha = a(\tilde{x})$: then $P_1\|P_2 \xrightarrow{a(\tilde{x})} P_1'\|P_2$ with $P_1 \xrightarrow{a(\tilde{x})} P_1'$ and $P_2 \xrightarrow{a:}$ . We can apply SYNC to mimic this transition.

$$\frac{[(\!|P_1|\!)_p] \xrightarrow{\{\}\triangleright\mathsf{tt}(a,\ \tilde{x})} [(\!|P_1'|\!)_p] \quad \dfrac{[(\!|P_2|\!)_p] \xmapsto{\{\}\triangleright\widetilde{\mathsf{tt}(a,\ \tilde{x})}} [(\!|P_2|\!)_p]}{[(\!|P_2|\!)_p] \xrightarrow{\{\}\triangleright\mathsf{tt}(a,\ \tilde{x})} [(\!|P_2|\!)_p]}}{[(\!|P_1|\!)_p] \ \| \ [(\!|P_2|\!)_p] \xrightarrow{\{\}\triangleright\mathsf{tt}(a,\ \tilde{x})} [(\!|P_1'|\!)_p] \ \| \ [(\!|P_2|\!)_p]}$$

Again we have that: $(\!|P_1'\|P_2|\!)_c \simeq [(\!|P_1'|\!)_p] \| \ [(\!|P_2|\!)_p]$.

3. if $\alpha = \bar{a}\tilde{x}$ then $P_1\|P_2 \xrightarrow{\bar{a}\tilde{x}} P_1'\|P_2$ with $P_1 \xrightarrow{\bar{a}\tilde{x}} P_1'$ and $P_2 \xrightarrow{a:}$. We can apply COML. There is also the symmetric case for rule COML.

$$\frac{[(\!|P_1|\!)_p] \xrightarrow{\{\}\triangleright\overline{\mathsf{tt}}(a,\ \tilde{x})} [(\!|P_1'|\!)_p] \quad \dfrac{[(\!|P_2|\!)_p] \xmapsto{\{\}\triangleright\widetilde{\mathsf{tt}(a,\ \tilde{x})}} [(\!|P_2|\!)_p]}{[(\!|P_2|\!)_p] \xrightarrow{\{\}\triangleright\mathsf{tt}(a,\ \tilde{x})} [(\!|P_2|\!)_p]}}{[(\!|P_1|\!)_p] \ \| \ [(\!|P_2|\!)_p] \xrightarrow{\{\}\triangleright\overline{\mathsf{tt}}(a,\ \tilde{x})} [(\!|P_1'|\!)_p] \ \| \ [(\!|P_2|\!)_p]}$$

□